\documentclass[manuscript]{emulateapj}

\slugcomment{Accepted on February 7th 2010 for publication in The Astrophysical Journal}
\shorttitle{Chemodynamics of Compact Stellar Systems in NGC 5128}
\shortauthors{Taylor et al.}

\begin{document}

\title{Chemodynamics of Compact Stellar Systems in NGC 5128: How similar are globular clusters, ultra-compact dwarfs, and dwarf galaxies?\altaffilmark{$\star$}}

\author{Matthew A. Taylor\altaffilmark{1,2}, Thomas H. Puzia\altaffilmark{1}, Gretchen L. Harris\altaffilmark{3}, William E. Harris\altaffilmark{4}, Markus Kissler-Patig\altaffilmark{5}, \& Michael Hilker\altaffilmark{5}}
\altaffiltext{$\star$}{Based on observations made with ESO Telescopes 
at the Paranal Observatories under program 69.D-0094 and 69.D-0196.}
\altaffiltext{1}{Herzberg Institute for Astrophysics, 5071 West Saanich Road, Victoria, BC, V9E 2E7, Canada, {\it puziat@nrc.ca}}
\altaffiltext{2}{University of Victoria, 3800 Finnerty Road, Victoria, BC, V8P 1A1, Canada,
{\it mataylor@uvic.ca}}
\altaffiltext{3}{Department of Physics and Astronomy, University of Waterloo, Waterloo, ON, L8S 4M1, Canada, {\it glharris@astro.uwaterloo.ca}}
\altaffiltext{4}{Department of Physics and Astronomy, McMaster University, Hamilton, ON, L8S 4M1, Canada, {\it harris@physics.mcmaster.ca}}
\altaffiltext{5}{European Southern Observatory, Karl-Schwarzschild-Str.~2, 85748 Garching, Germany, {\it mkissler@eso.org, mhilker@eso.org}}

\begin{abstract}
Velocity dispersion measurements are presented for several of the most
luminous globular clusters (GCs) in NGC 5128 (Centaurus A) derived from
high-resolution spectra obtained with the UVES echelle spectrograph on the
8.2m {\it ESO/Very Large Telescope}.~The measurements are made utilizing a
penalized pixel fitting method that parametrically recovers line-of-sight
velocity dispersions (LOSVD). Combining the measured velocity dispersions
with surface photometry and structural parameter data from the
\emph{Hubble Space Telescope} enables both dynamical masses and
mass-to-light ratios to be derived. The properties of these massive
stellar systems are similar to those of both massive GCs contained
within the Local Group and nuclear star clusters and ultra-compact dwarf
galaxies (UCDs).~The fundamental plane relations of these clusters are
investigated in order to fill the apparent gap between the relations of
Local Group GCs and more massive early-type galaxies.~It is found
that the properties of these massive stellar systems match those of
nuclear clusters in dwarf elliptical galaxies and UCDs better than those
of Local Group GCs, and that all objects share similarly old
($\ga\!8$ Gyr) ages, suggesting a possible link between the formation and
evolution of dE,Ns, UCDs and massive GCs.~We find a very steep
correlation between dynamical mass-to-light ratio and dynamical mass of
the form $\Upsilon_{V}^{\rm dyn}\!\propto\!{\cal M}_{\rm
dyn}^{0.24\pm0.02}$ above ${\cal M}_{\rm dyn}\!\approx\!2\times10^6\,
M_\odot$.~Formation scenarios are investigated with a chemical abundance
analysis using absorption line strengths calibrated to the Lick/IDS index
system.~The results lend support to two scenarios contained within a
single general formation scheme.~Old, massive, super-solar [$\alpha$/Fe]
systems are formed on short ($\la100$ Myr) timescales through the merging
of single-collapse GCs which themselves are formed within single,
giant molecular clouds.~More intermediate- and old-aged ($\sim\!3\!-\!10$
Gyr), solar- to sub-solar [$\alpha$/Fe] systems are formed on much longer
($\sim$Gyr) timescales through the stripping of dE,Ns in the $10^{13}
\!-\! 10^{15} M_\odot$ potential wells of massive galaxies and galaxy
clusters.
\end{abstract}


\section{Introduction}
\label{ln:intro}
In the past decade, studies of the central regions of the Virgo and
Fornax galaxy clusters have revealed dozens of massive compact objects
in the luminosity range $-13.5\!\la\! M_{V}\!\la\!-11.5$ mag, with
half-light radii between 10 and 30 pc and central velocity dispersions
$25\!\la\!\sigma_{o}\!\la\!45$ km~s$^{-1}$ \citep{hil99, dri00, mie02,
mie04, has05, jon06, hil09}.~\cite{phi01} dubbed the new objects
``ultra-compact dwarf'' galaxies (UCDs), a term that was immediately
controversial because of the galaxian origin that it implies. The term
has since come to describe a ``mixed bag of objects", in the words of
\cite{hil09}, defined only by observational parameters and saying
nothing about evolutionary history. When combined with other structural
parameters, these objects show similarities to both massive globular
clusters, and nuclear star clusters in dwarf elliptical galaxies
\citep{cot06}, raising the possibility of a shared formation history
among some of these objects. The GC system of the nearby giant
elliptical galaxy NGC 5128 offers an interesting target to compare
the properties of massive stellar systems to those of typically less
massive systems contained within the Local Group. The relatively simple
structures of GCs are fairly well approximated by isotropic, single-mass
King (1966) models and the structural parameters of these stellar
systems such as core- and half-light radii, central concentrations,
surface brightnesses, velocity dispersions, mass-to-light ratios, etc.,
have been shown to inhabit only a narrow region called the fundamental
plane (FP; \citealt{djo95}).

While the FP relations are well established for Milky Way GCs, there
have been only limited studies done for the properties of GCs in
external galaxies. The external galaxy for which similar studies have
mostly been done is M31, a disc galaxy similar to the Milky Way. The
structures of its GCs appear to share similar properties to those in
the Milky Way \citep{fus94,hol97,bar02,bar07}. \cite{djo97} undertook one
such study and report velocity dispersions for four of the M31 clusters
being as high as $\sigma$ $\gtrsim$ 20 kms$^{-1}$, implying masses 
of the order of the largest galactic globular cluster $\omega$Cen
\citep{mey95, soll09}.~It is interesting to note that both $\omega$Cen and
G1, the largest of the four M31 clusters, share similar peculiarities.
$\omega$Cen is the most flattened galactic globular cluster \citep{whi87}
with significant [Fe/H] differences among its member stars
\citep{nor95,pan02}, two characteristics shared by G1, although it should
be noted that G1's metallicity spread was inferred from the photometric
width of the RGB \citep{mey01}.

A possible explanation for the odd properties of these massive clusters is
that they could be the nuclei of tidally stripped dwarf elliptical
galaxies \citep[dEs;][]{zinn88, hil00,mey01,gne02,bek03}.~This idea seems
to be supported by studies of massive, compact stellar systems in the
Virgo \citep[DGTOs;][]{has07}, and Fornax \citep[UCDs;][]{dri03} galaxy
clusters, which appear to share characteristics with both nuclear
star clusters in dEs (dE,Ns) and the most massive globular clusters in M31
and the MW. The Virgo and Fornax DGTO/UCDs show a size-luminosity relation
similar to that of dE/dE,Ns while having a range of sizes similar to that
of massive local GCs \citep{evs08,hil09}.

Another explanation is that massive compact systems could be merged young
massive clusters (YMCs) produced by violent galaxy-galaxy mergers
\citep{fel02}. \cite{mie06} present [Fe/H] estimates from a spectroscopic
study of 26 compact objects in the central regions of Fornax which were
compared to 5 known Fornax dE,Ns. It was found that the mean metallicities
of compact objects with $M_V\!<\!-11.0$ mag are $\sim\!0.8$ dex higher
than for the dE,Ns while their $V\!-\!I$ colors suggest that they are
generally younger (and/or more metal-poor) than the dwarf nuclei. The mean
metallicities agreed much more with those for YMCs of comparable masses,
thus consistent with the YMC merging formation scenario. Studies of massive
star clusters discovered in the Virgo cluster \citep{has05,jon06} reveal
brighter and bluer objects compared to Fornax compact objects and are
suggestive of the stripping scenario. These results together imply that
both formation scenarios are possible, perhaps even likely, but that what
occurs could be dependent on environmental factors such as local galaxy
density.

To make progress on the question of the origin of compact stellar systems
it is necessary to study clusters that bridge the mass gap between the
systems like $\omega$Cen and G1, and dwarf galaxies in a variety of
environments. Being the central galaxy in a large group, NGC 5128 not only
provides a less dense environment compared to the central regions of
galaxy clusters, but evidence for recent merger activity in the form of
the dust lane, faint shells \citep{mal78}, and a young tidal stream
\citep{pen02} make it a likely environment for YMCs to be formed. NGC~5128
is also the nearest giant galaxy to the Milky Way beyond M31 and thus it
is possible to get high resolution spectra and structural parameters of
its large population of $\sim\!2000$ star clusters, many of which are at
the upper end of the globular cluster mass distribution
\citep{har84,har02}. All of these features make NGC 5128 an interesting
target for such a study.

The goal of this work is to study correlations between dynamical
properties, such as dynamical masses and mass-to-light ratios, and basic
evolutionary parameters, e.g.~relative ages, metallicities, and
[$\alpha$/Fe] ratios. The ratio of $\alpha$-capture to Fe-peak elements
[$\alpha$/Fe]) can provide information on the star formation history (SFH)
of stellar populations. Super-solar [$\alpha$/Fe] is indicative of a short
burst of SF, typical of GCs and elliptical galaxies, while sub-solar
values mean a more quiescient or drawn out SFH. Ages, metallicities, and
[$\alpha$/Fe] ratios are usually estimated from integrated-light spectra
by the measurement of absorption-line features that have been calibrated
to the Lick/IDS index system \citep{bur84,wor94,wor97,tra98} and compared
to models that assume simple stellar populations \citep[SSPs; see
e.g.][]{tho03}. These parameters then provide useful clues to the
evolutionary pasts of the objects.

In this work we measure dynamical masses of compact star clusters in NGC
5128 using a homogeneous analysis method based on the penalized pixel
fitting (pPXF) code of \cite{cap04}. The results compare favourably with
those of \cite{rej07}, who derive dynamical masses for 27 objects based on
$\sigma$ measured using a cross-correlation technique.~These dynamical
masses are then combined with photometric structural parameters to derive
the FP relations and we compare them to those of objects ranging in mass
from the smallest Local Group GCs to giant elliptical galaxies. Chemical
abundances from the literature for the NGC 5128 objects derived using the
Lick/IDS system are used to investigate the ages and star formation
histories. These properties, along with dynamical masses and mass-to-light
ratios, are directly compared to those for early-type galaxies, UCDs and
Local Group GCs.

The paper is organised as follows. \S2 summarizes the data and
observations.~The measurement of $\sigma$ is described in \S3 as well as
the calculations that lead directly to dynamical mass estimates (${\cal
M}_{\rm dyn}$) and both photometric and dynamical mass-to-light ratios
($\Upsilon_{V}^{\rm phot}$ and $\Upsilon_{V}^{\rm dyn}$).~\S4 compares the
NGC 5128 clusters to other compact stellar systems using a comparison of
key structural parameters, a fundamental plane and $\kappa$-space analysis
as well as a comparison of stellar population properties. We conclude with
a discussion of the main results in light of the various formation
scenarios in \S5.

\section{Observations}
\subsection{Instrumental Setup}

High resolution spectra of GCs in NGC~5128 were obtained with the
UVES spectrograph at the 8.2m Very Large Telescope on Cerro Paranal in
Chile.~UVES is a two-arm, cross-dispersed echelle spectrograph operating
at the Nasmyth B focus of UT2 (Kueyen).~The two arms of UVES cover the
wavelength ranges 3250--4500 \AA\ and 4700--6800 \AA\ for the blue and red
arms respectively. The spectrograph was set to its Dichroic \#1 mode for
the observations taken in program  69.D-0094 in order to obtain spectra on
both arms simultaneously, whereas the spectra obtained in program
69.D-0169 were taken with the red arm only. With the 1\arcsec \ slit used
for all observations, UVES provides a spectral resolution of
$R\!\simeq\!40000$, corresponding to a velocity resolution FWHM of
7.5 km s$^{-1}$ ($\sigma\!\simeq\!3.2$ km s$^{-1}$).~For further details
of the UVES operating specifications we refer to \cite{dek00}.

\subsection{Cluster Sample}

Pipeline-reduced high-resolution spectra for 23 globular clusters were
obtained from the ESO/UVES archive and are summarized in
Table~\ref{tab:gcobs}, which lists cluster identifications according to
\cite{har02} (old) along with the newer, homogeneous identifiers of
\cite{woo07} (new), the total integration times, apparent $V$-band
magnitudes, and the average S/N over the wavelength ranges used. The total
integration times listed are sums of three or four individual
sub-exposures taken during the observations. The apparent $V$ magnitudes
listed are taken from the \cite{pen04} catalogue where available;
otherwise, the magnitudes are from the original discovery publications. The
airmass was in the range 1.054--1.986 for all cluster observations and the
seeing ranged from 0.52\arcsec to 1.35\arcsec.

\subsection{Template Stars}

All template stars were observed with the same instrumental setup as the
clusters, but with only single exposures of a few seconds each.~The
template stars cover a spectral range of G0--M0.5 and luminosity
classes Iab, Ib, IIa, II, and III as listed in Table~\ref{tab:stars}.~The
seeing and airmass during the collection of the template spectra were
comparable to the cluster observations. All template stars were observed around
April 2002 in two separate observing programs with some overlap in their
target lists.~To increase the S/N, any of the template stars for which
multiple observations were made were summed together. Because of the
heliocentric $v_{r}$ corrections applied to the spectra prior to their
combination a systematic uncertainty of not more than $\sim\!0.5$ km
s$^{-1}$ is introduced to the final velocity dispersion measurement. This
is added in quadrature to the total error budget resulting from the pPXF
analysis  (see Sect.~\ref{ln:losvd}) using the corresponding template
star.

\subsection{Data Processing}

The basic data reduction was done by the ESO-UVES pipeline, which
provided flux- and wavelength-calibrated, de-biased spectra that were
further processed using a combination of IRAF\footnote{IRAF is
distributed by the National Optical Astronomy Observatory, which is
operated by the Association of Universities for Research in Astronomy,
Inc., under cooperative agreement with the National Science Foundation.}
tasks and IDL programs.~The signal-to-noise ratios (S/N) per pixel for
all of the clusters and template stars were measured over the wavelength
ranges described in \S2 using the \emph{splot} task in IRAF and are
listed in Tables~\ref{tab:gcobs} and \ref{tab:stars}. The S/N over the
REDL($\sim$4800--5750\AA) and REDU($\sim$5800--6800\AA) regions of the
spectra were high enough to perform the convolutions for the subsequent
velocity dispersion analysis.~It was found that the S/N over the BLUE
arm ($\sim$3300--4500\AA) was too low for accurate $ \sigma$
measurements with errors well above the expected velocity dispersion of
our target clusters. For this reason the $\sigma$ values measured from
the BLUE CCD are dropped from the subsequent analysis.

\section{Analysis}

\subsection{Penalized Pixel Fitting}
We use the penalized pixel-fitting code (pPXF) described in detail in
\cite{cap04} to measure the line-of-sight velocity dispersion (LOSVD),
$\sigma$, of our target GCs. In summary, pPXF expands the LOSVD as a
Gauss-Hermite series of the form
\begin{equation}
L(v)=\frac{e^{-(1/2)y^{2}}}{\sigma\sqrt{2\pi}}[1+\sum_{m=3}^{M}{h_{m}H_{m}(y)}]
\end{equation}
where $y=(v-v_{r})/\sigma$ and $H_{m}$ is the $m$-th Hermite polynomial.
pPXF then makes use of an input `initial guess' for the relative radial
velocity, $v_{r}$, of the template star as well as an estimated $\sigma$
for the cluster in order to calculate the best fitting $v_{r}$, $\sigma$
and the Gauss-Hermite moments $h_3, ..., h_M$ for the best fitting LOSVD.
The code uses a penalty function that is derived from the integrated
square deviation of the line profile from the best-fitting Gaussian. The
fit is iterated until the perturbed penalty function cannot decrease the
variance of the fit any further. This key feature of pPXF serves to bias
the solution towards a Gaussian shape when the S/N is low but allows the
solution to reproduce the higher-order details of spectral features when
the S/N is high. Features of note, that are implemented in the pPXF
routine, include fitting of an optimal template together with the
kinematics to minimize the impact of template mismatches as well as an
iterative sigma clipping routine that acts to clean the spectra of any
residual bad pixels or cosmic rays \citep[see][for details]{cap04}.

\begin{figure}
\plottwo{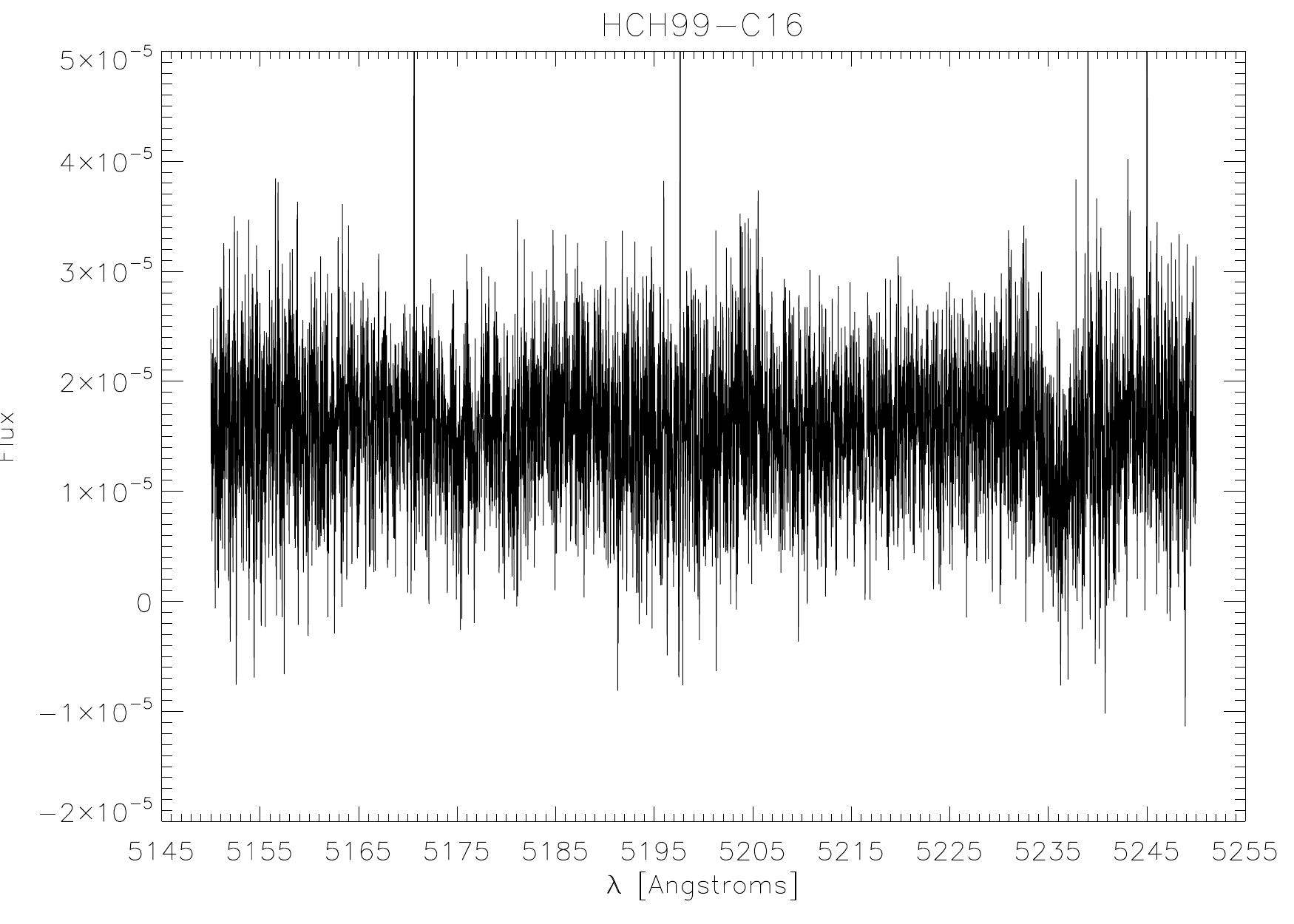}{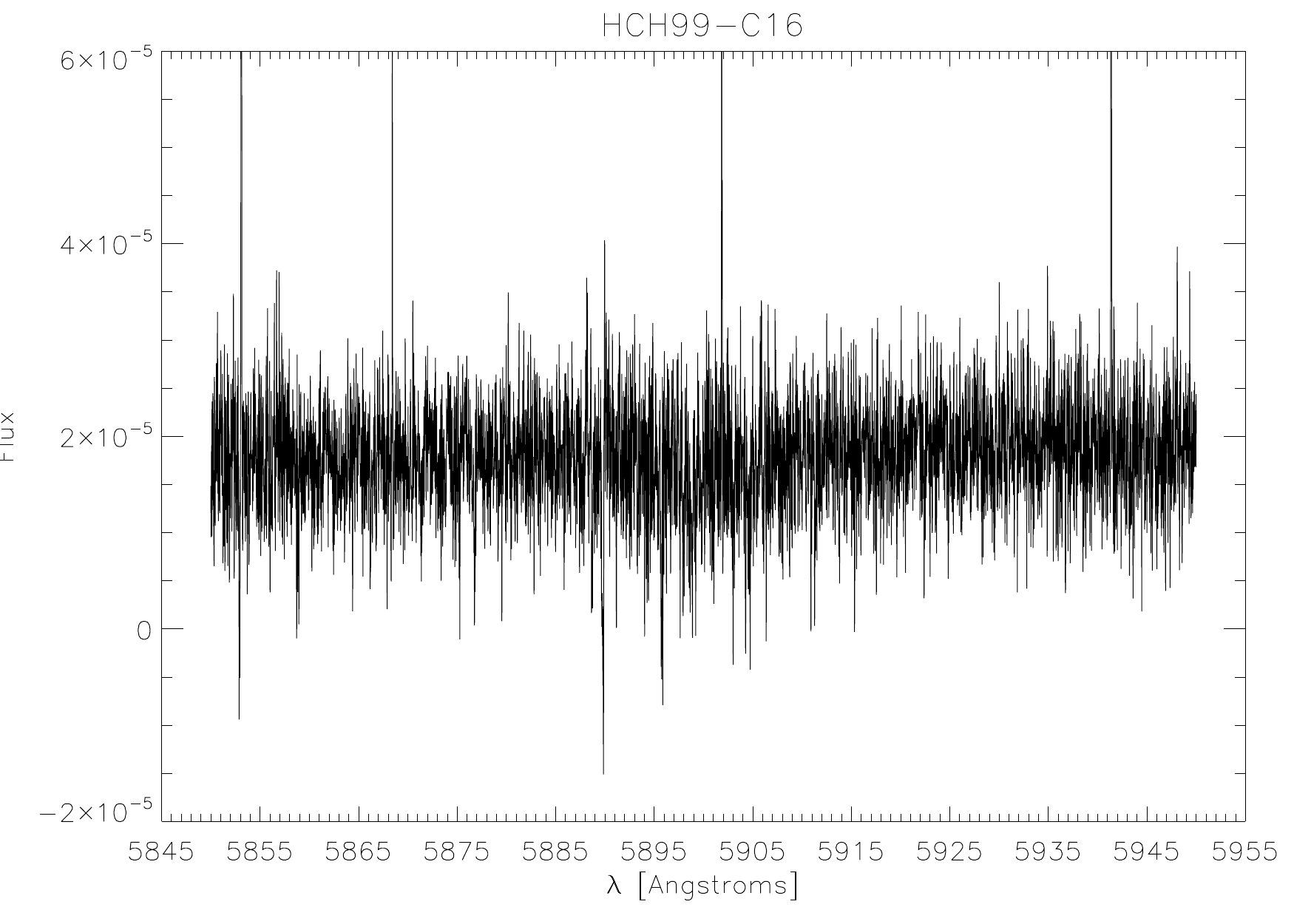}
\plottwo{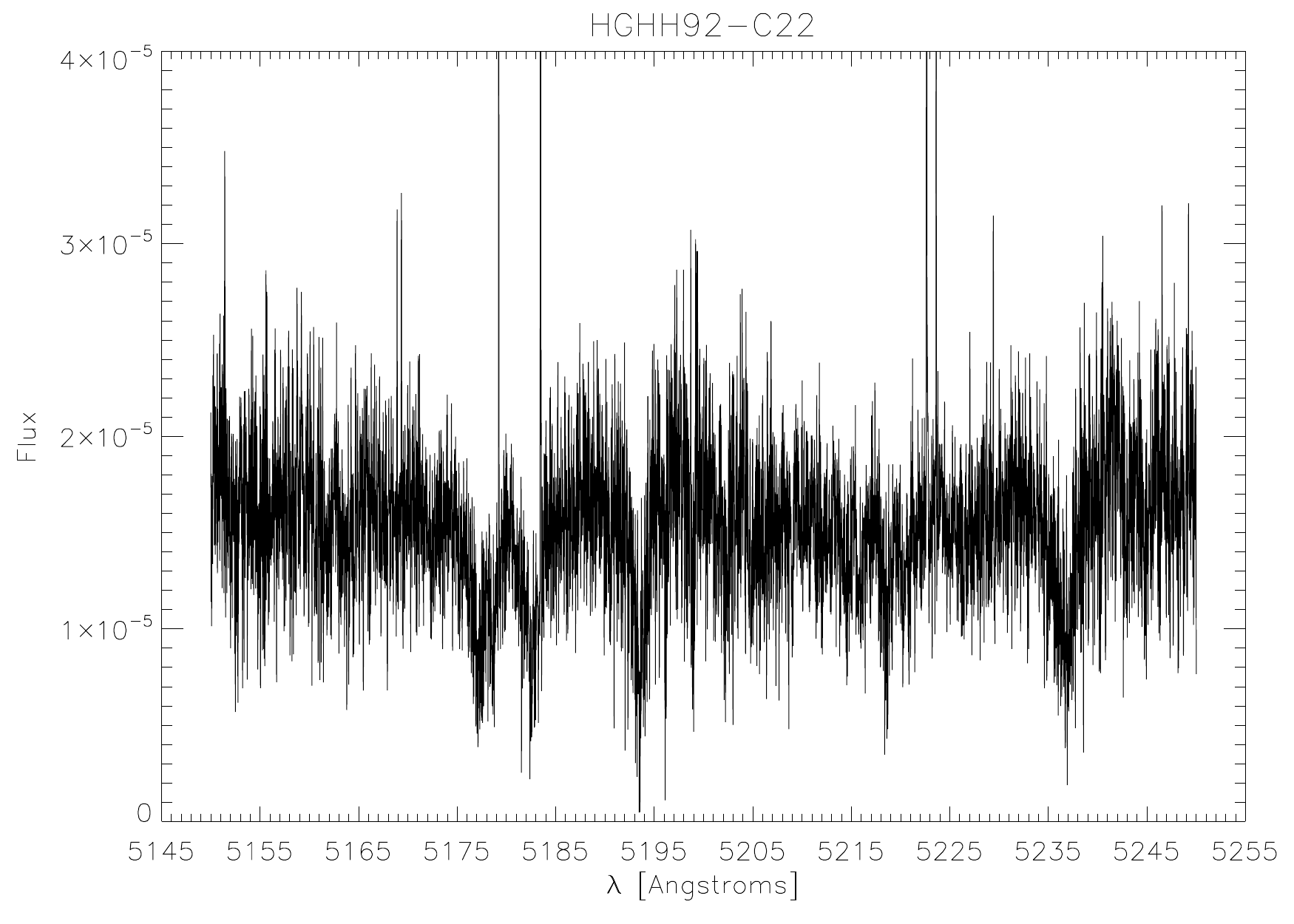}{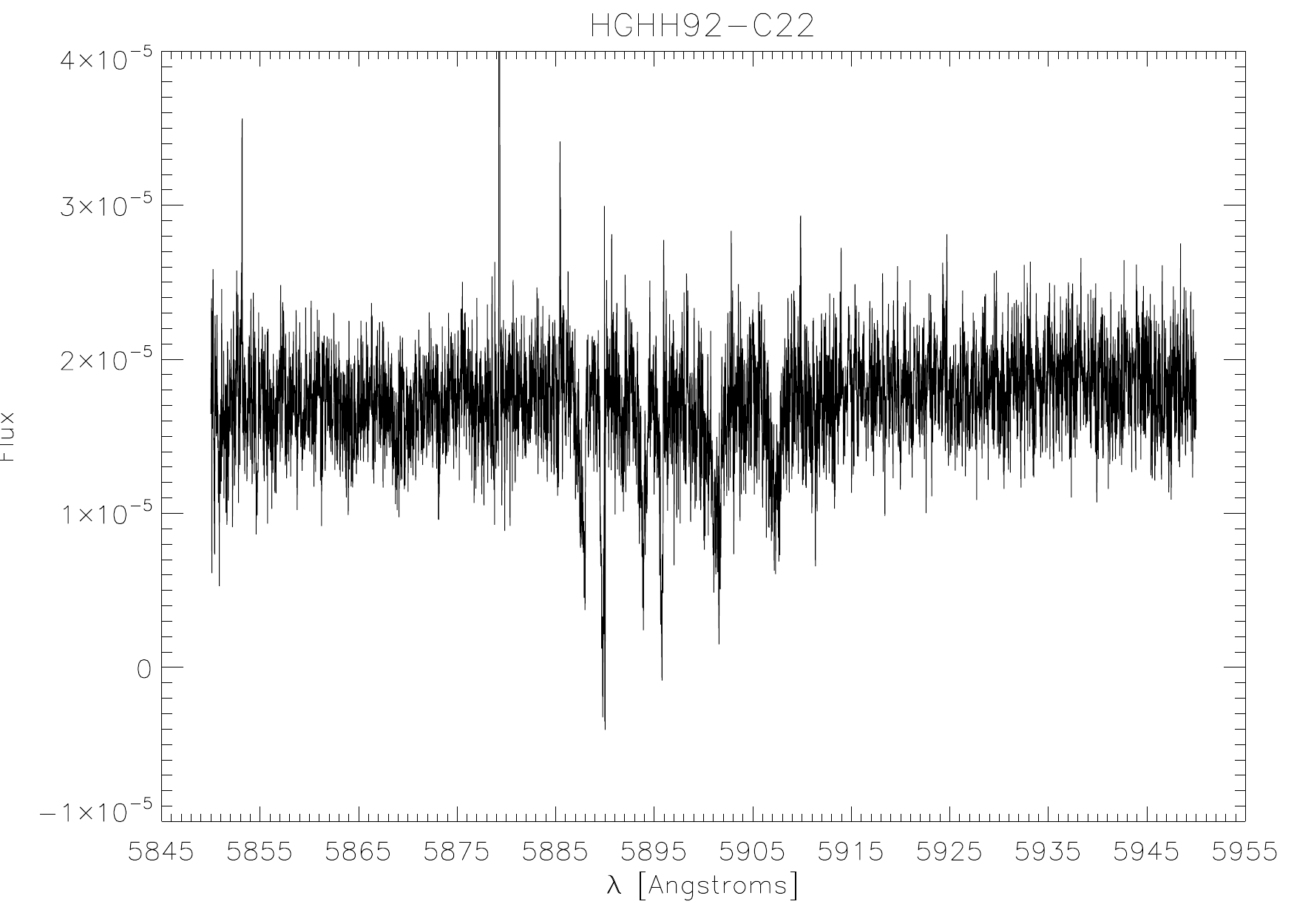}
\plottwo{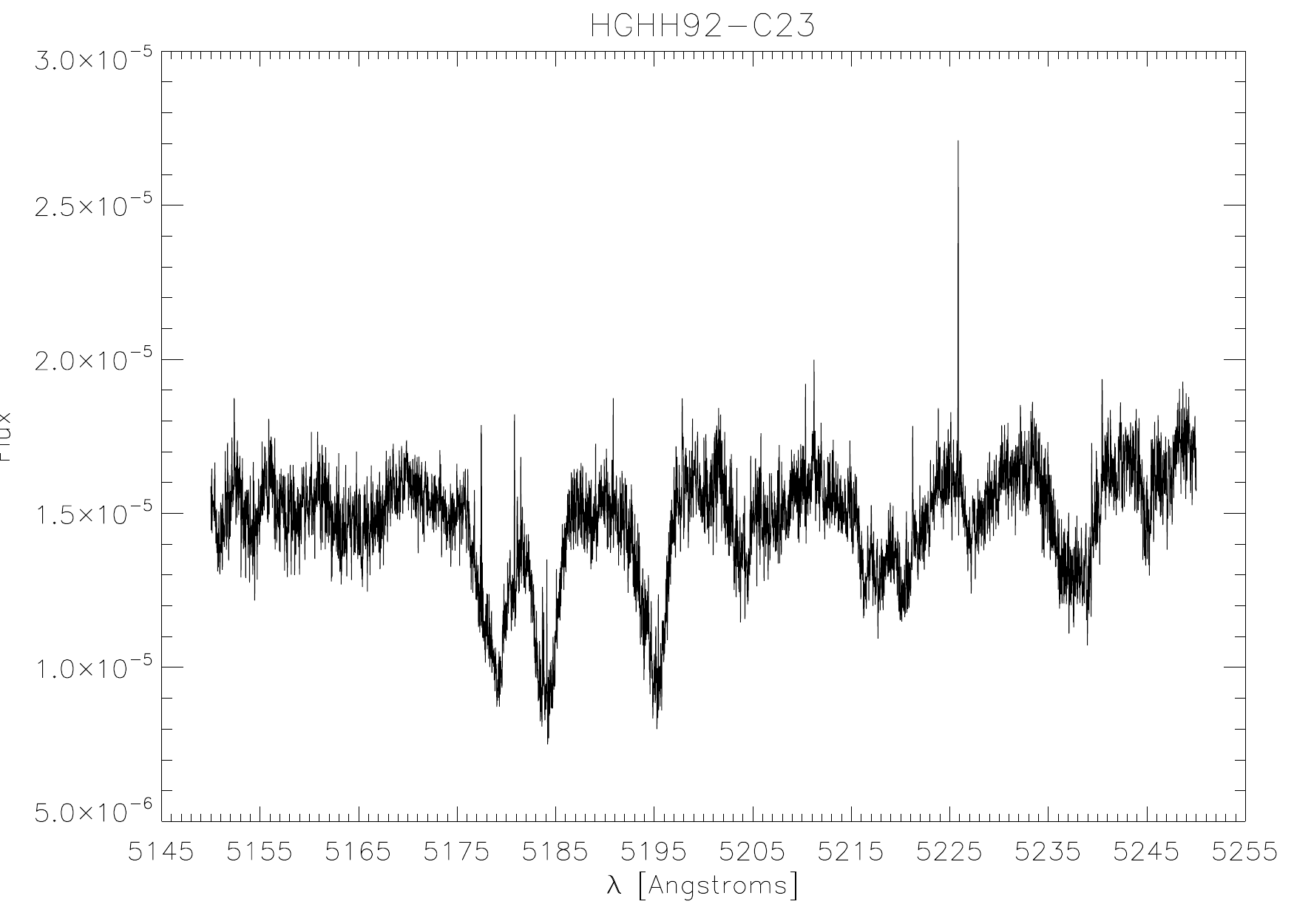}{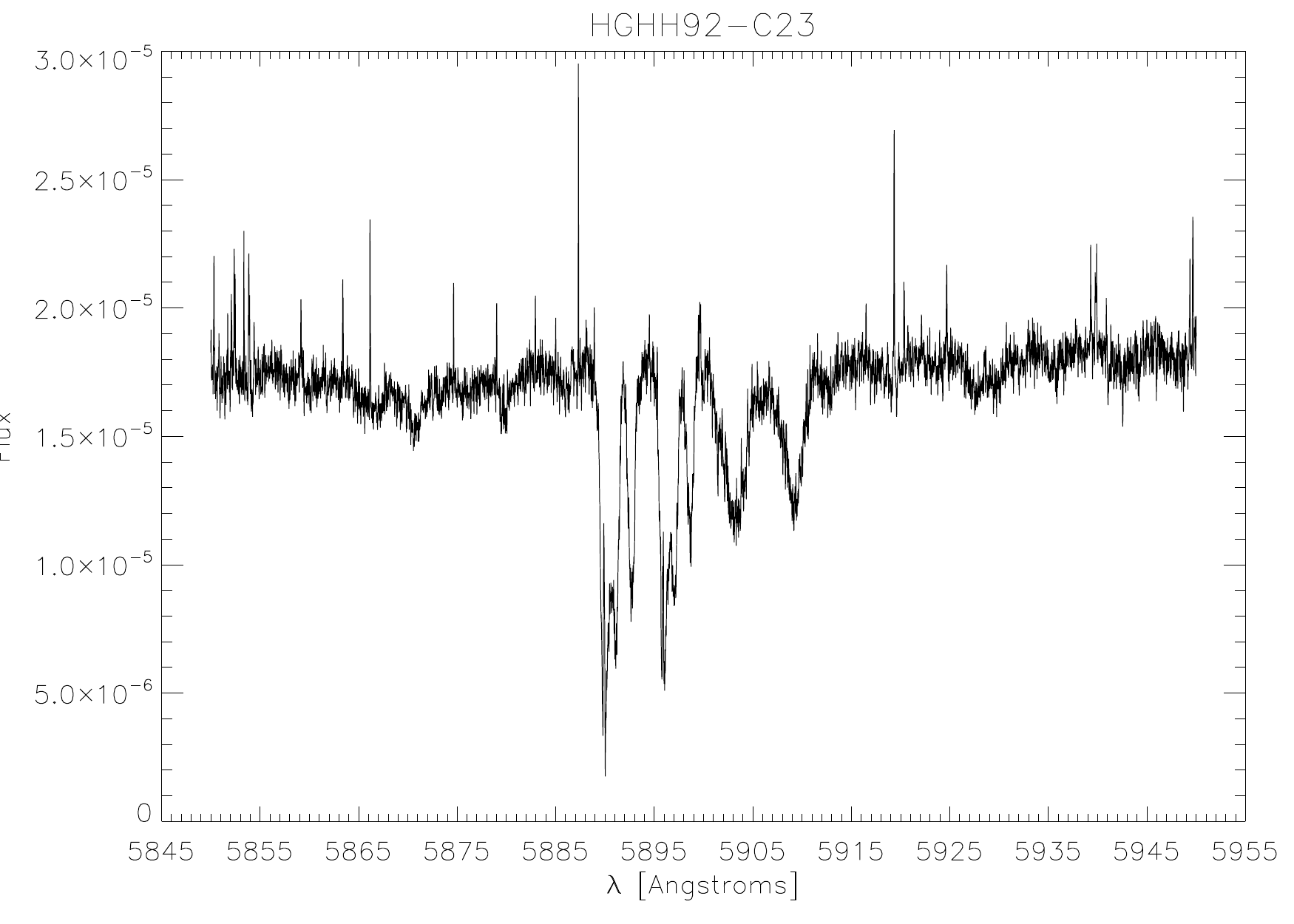}
\caption{Spectral regions around the Mg$b$ triplet (left panels) and several Fe {\sc i}, Na {\sc i}, Si {\sc i}, and TiO features (right panels) used by pPXF for the convolutions contained within the constrained REDL and REDU ranges, normalized by total flux.}
\label{fig:spec}
\end{figure}

In order to maximize the accuracy of the LOSVD measurements the
wavelength range of the object spectra was constrained to $\sim$500--600 \AA\
sections around prominent features.~Extensive experimentation showed that
the wavelength regions which provided the most robust results from the
REDL and REDU CCDs were 4951--5550\AA\, containing the Mg\emph{b} triplet
and several Fe features, and 5850--6400\AA, containing the NaD doublet and
two TiO molecular bands. Figure~\ref{fig:spec} shows a representative
sample of our cluster spectra constrained to these regions for both REDL
and REDU ranging from the lowest S/N spectra at the top to the highest S/N
at the bottom. The spectra were left with enough continuum for pPXF to
perform the convolutions up to $\sigma\!\approx\!100$ km s$^{-1}$.

In order to minimize the input scatter, one of the parameters required by
pPXF, $v_{r}$, was measured using the \emph{fxcor} task in IRAF for all of
the clusters against each of the template stars. The averaged results,
which are in good agreement with those in the literature
\citep[e.g][]{rej07,woo09}, are listed in Table~\ref{tbl:sigcomp} along
with the standard errors. The measured $v_{r}$ from \emph{fxcor} was
obtained with Welch filtering using cut-on and cut-off wavenumbers of 0
and 500, respectively, in the constrained regions. Adjusting the filtering
parameters can produce small variations in the measured $v_{r}$, which
might add to uncertainties in the final $\sigma$. To test this, pPXF was
used to measure $\sigma$ with the input $\sigma$ held constant as the
input $v_{r}$ was varied for each cluster against the templates. The
results of this test showed that the scatter in the measured $\sigma$
values was no more than 1 km s$^{-1}$, indicating that pPXF is relatively
insensitive to variations in $v_{r}$ assuming that the input $v_{r}$ is a
`good' guess within a few tens of km s$^{-1}$. Therefore, any uncertainty
in measuring the `initial guess' $v_{r}$ with \emph{fxcor} should not
translate into larger uncertainties of the final $\sigma$ measurement.

\subsection{LOSVD Measurement}
\label{ln:losvd}
The pPXF code does indicate a sensitivity to very high or low input
estimates for $\sigma$, so before any $\sigma$ measurements were made, the
input $\sigma$ was varied between 5--50 km s$^{-1}$ in steps of 5 km
s$^{-1}$ while $v_{r}$ was held constant at the measured values. This
range was selected because the typical $\sigma$ of massive clusters in the
Local Group is of the order of $\la 30$ km s$^{-1}$ \citep{mho04}.
For this reason, all of the input $\sigma$ estimates are `good' guesses,
and account for any under- or over-estimates of the cluster velocity
dispersions. The resulting measurements from all template star-GC
combinations were $\kappa\!-\!\sigma$ clipped to within one standard
deviation based on measurements from all 21 templates, which includes any
$v_{r}$ uncertainties from \emph{fxcor}, the stellar template library, and
systematic uncertainties intrinsic to the pPXF code due to template
mismatches. For each cluster, the mean of the clipped $\sigma$ was
calculated and recorded for both REDU and REDL spectral regions. The
errors were taken as one standard deviation of the mean-clipped $\sigma$
measurements.

\begin{figure}
\plotone{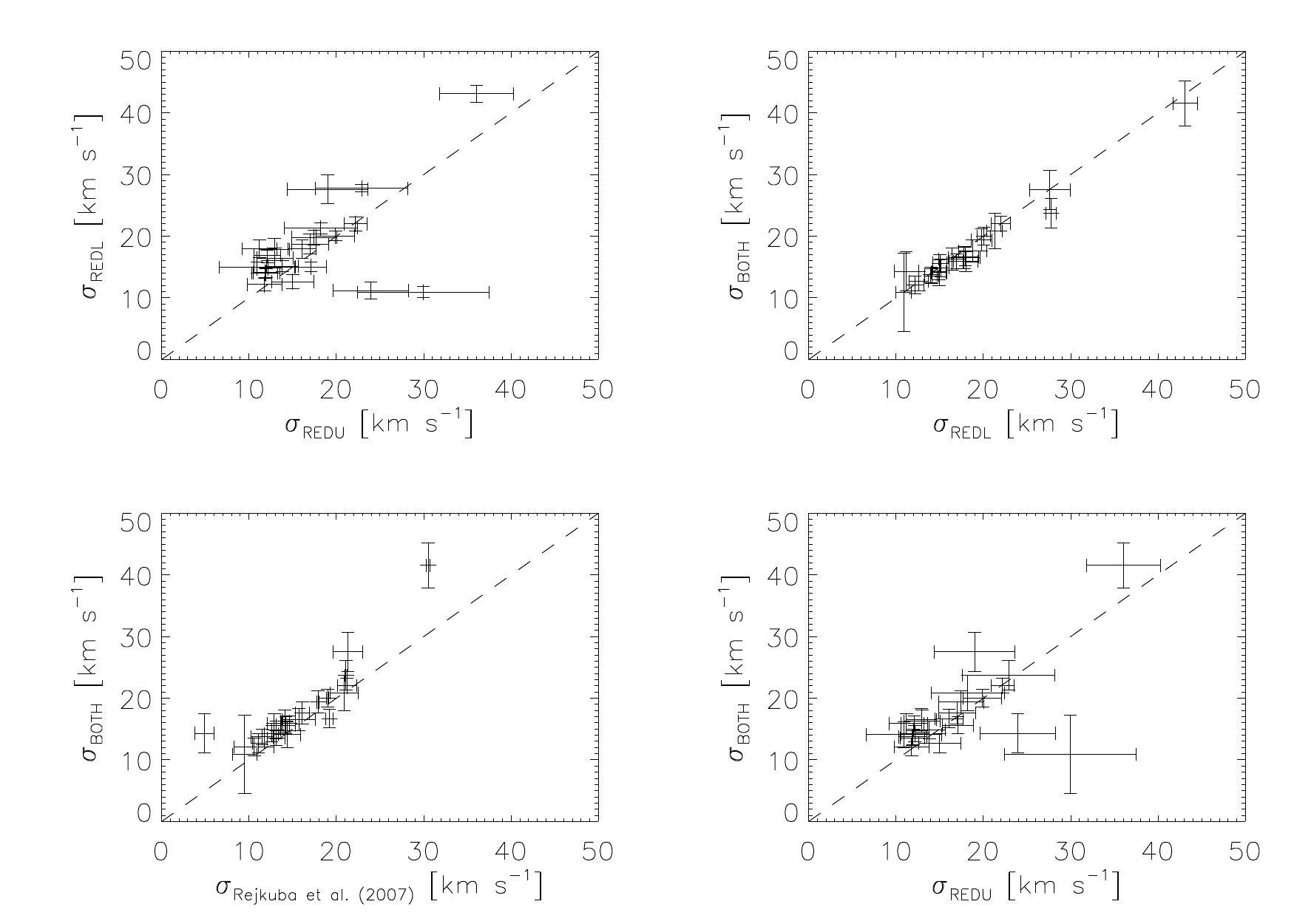}
\caption{Comparison of various LOSVD measurements performed with pPXF and from the literature. The upper left panel shows $\sigma$ measured from both REDL and REDU CCDs. The rest of the panels compare $\sigma$ from the combined REDL and REDU measurements against only the REDL CCD (upper right), REDU CCD (lower right), and the measurements recorded by \cite{rej07} (lower left). In each panel, the unity relation is shown by the dashed line.}
\label{fig:sigs}
\end{figure}

Due to the strength and distinctiveness of the Mg\emph{b} triplet in the
REDL spectral range compared to the many, but less distinct, features in
the REDU range (5850--6400 \AA), the measured $\sigma$ has much more
robust results from the REDL CCD than from the REDU CCD.
Figure~\ref{fig:sigs} (top and lower right panels) shows a comparison of
the $\sigma$ measurement from both CCDs combined against those from only
the REDL or REDU CCD, and against each other. The plots indicate that
almost all of the results from both CCDs fall within errors from either
of the single CCDs, thus justifying the use of the weighted mean from
both CCDs, as listed in Table~\ref{tab:ds}. As as check against the
possibility of interstellar absorption affecting the $\sigma$
measurements around the NaD regions of the REDU spectra, measurements
were made for the representative spectra shown in Fig.~\ref{fig:spec} as
well as the outlier GC0382 (see Fig.~\ref{fig:sigs} both with and
without these features. The results suggest a negligible effect for low
S/N spectra; however, for higher S/N, $\sigma$ could be over-estimated
by $\sim\!1\!-\!2$ km s$^{-1}$, indicating that the larger error bars
for the REDU measurements could be at least in part due to interstellar
absorption.

The range of $\sigma$ measured from our sample was
$10.90\leq\sigma\leq41.53$ km s$^{-1}$ with the former measured for
GC0217 and the latter for GC0330. Figure~\ref{fig:sigs} (lower left
panel) shows the pPXF results against the $\sigma$ recorded by
\cite{rej07}, which were obtained using a cross-correlation technique
and whose measurements were shown to agree well with those of
\cite{mho04}. The measurements agree very well for 21 out of the 23
clusters, although we measure significantly higher $\sigma$ for the two
clusters GC0330 and particularly GC0382. \cite{rej07} reported
$\sigma$=4.9 km s$^{-1}$ for GC0382, but this was based on a CCF with
multiple peaks due to additional absorption features in the spectrum.
They suggest that these abnormalities could be due to spatially
unresolved Galactic foreground stars. While we do not offer an
explanation for this discrepancy, considering the restricted spectral
ranges described above and the fact that pPXF has a certain amount of
robustness against features such as these, we tend to prefer our
measurements over previous determinations from the literature, but bear
in mind the significant confusion regarding this object.

\subsection{Structural Parameters}
\label{ln:sp}
Structural parameters, such as projected core-radius, $r_c$, the King
concentration parameter, c, and the projected half-light radius,
R$_{h}$, are important for aperture corrections to our LOSVD
measurements and are listed in Table~\ref{tab:ds}. For all but the
clusters from the \cite{hol99} study (with HCH99 designations, see
Tbl.~\ref{tab:gcobs}), R$_{h}$ and r$_{c}$ were quoted from \cite{har02}
and the HCH99 cluster parameters were quoted from their paper. In
converting R$_{h}$ and r$_{c}$ to pc, a distance modulus of
$(m\!-\!M)_{0}$=27.88$\pm$0.05 mag, corresponding to a distance of
3.8$\pm$0.1 Mpc, was used for NGC 5128 \citep{har09}.

\subsection{Aperture Effects}
\label{ln:ap}
The central velocity dispersion, $\sigma_{o}$, plays a pivotal role in
dynamical mass estimates.~However, LOSVD measurements in the literature
are often performed at various radii depending on the instrumental setup
and target sample of the particular study. Because of varying seeing
conditions, target distances, and variations in the instrumental
specifications, such as spectral and spatial resolution and sampling, the
light entering a given aperture may not be representative of the object
being observed as a whole and thus the true central velocity dispersion is
often inaccessible and must be extrapolated using assumptions on stellar
orbit distributions.~\cite{mie08} show that aperture corrections for
objects in the mass range of massive GCs or UCDs can be significant, and
so appropriate corrections were applied to the NGC 5128 objects.

The aperture corrections applied to our data were performed using the code
discussed in detail in \cite{hil07} and \cite{mie08}, which uses the basic
cluster structural parameters that define the shape of the cluster light
profile (i.e.~R$_h$ and $c$, see Table~\ref{tab:ds}). To create a
3-dimensional stellar density profile, a King surface density profile is
assumed, from which the cumulated mass function M($<\!r$), potential
energy $\phi (r)$, and energy distribution function $f$(E) are computed.
The 3-D profile and $f$(E) are then used to create an N-body
representation of the cluster; in our case, $10^5$ particles were
simulated for each cluster. Each particle is then convolved with a
Gaussian corresponding to the seeing during the observations. A light
profile is then generated and used to calculate the velocity dispersion at
a given radius or within a given aperture.

To estimate the uncertainties of the inferred central velocity
dispersions, each of the clusters were modeled three times. Firstly, the
code was run using ${\rm R}_{h}$, $c$, and $\sigma_{\rm ppxf}$ from
Table~\ref{tab:ds} to get $\sigma_{o}$. Maximum and minimum values for
$\sigma_{\rm ppxf}$, ${\rm R}_{h}$ and $c$ were calculated by adding or
subtracting the errors listed in Table~\ref{tab:ds} which were used to
determine the upper and lower uncertainties for $\sigma_{o}$ by
re-modelling the clusters using $\sigma_{\rm ppxf,max}$, ${\rm R}_{h,min}$
and $c_{max}$ (for $\sigma_{o,max}$) and $\sigma_{\rm ppxf,min}$, ${\rm
R}_{h,max}$ and $c_{min}$ (for $\sigma_{o,min}$). The differences between
$\sigma_{o}$ and $\sigma_{o,max/min}$ were then assigned as the upper and
lower uncertainties which were then propagated through Eq.~\ref{eq:mdyn}
to calculate the absolute errors on ${\cal M}_{\rm dyn}$.

The first two columns of Table~\ref{tbl:sigcomp} show the velocity
dispersions measured in our slit apertures, $\sigma_{\rm ppxf}$, against
those calculated by the models with the cluster structural parameters,
$\sigma_{\rm m,obs}$, as a demonstration of how well the model reproduces
the observed values. We use the model to correct our $\sigma_{\rm ppxf}$
measurements to the central velocity dispersion, $\sigma_{o}$, as listed
in the last column of Table~\ref{tbl:sigcomp}. The third column of
Table~\ref{tbl:sigcomp} shows $\sigma_{m,global}$, which is the $\sigma$
predicted by the model if an aperture is used that covers the cluster to
the tidal radius.

\subsection{Aperture Corrections of Literature Data}
\label{ln:ap_lit}
Our goal is to compare our sample GCs to other stellar systems.~For that
purpose we searched the literature for central velocity dispersion
measurements. \citet[and references therein]{str09} provide $\sigma_o$
measurements for a number of M31 clusters based on high-resolution
spectra, all of which have been aperture corrected to the cluster
centers using the corrections described in \cite{bar07}.~\cite{geh03}
list data for six dEs (five of which are nucleated), which they analyze
in two ways. First they model the objects as dEs, fitting a S\'{e}rsic
model to the surface brightness profile for $r\!>\!1$\arcsec, then as
dE,Ns by fitting a Plummer model to the surface brightness profile
within $r\!<\! 1$\arcsec.~While their profile fitting method is not
completely consistent with ours, we use their measurements to
distinguish, at least, between the core and envelope parameters of
nucleated dwarf ellipticals.~In the following we break their $\sigma$
measurements into three groups: one non-nucleated dE, five dE,Ns (core
plus envelope), and only the cores of the same five dE,Ns. \cite{geh02}
define $\sigma_{o}$ of the dE,Ns as the velocity dispersion of the
nuclei (r $\!<\!$ 1\arcsec), and so in order to analyze the three groups
consistently we treat their $\sigma$ measurements all as observed
values, which are probably lower limits to $\sigma_{o}$.

Velocity dispersion measurements for the UCDs come from three
different sources. Data for 15 UCDs in the Fornax cluster and 11 in the
Virgo cluster were taken from \cite{mie08} and Patrick C\^{o}t\'{e} (private
communication) respectively, who both provide aperture corrected
$\sigma_{o}$ using the code of \cite{hil07}.~An additional six Virgo UCDs
were culled from \cite{evs07} who list $\sigma_{o}$ calculated using a
generalized King model, with the exception of one object with a King +
S\'{e}rsic profile, and aperture corrected to $r\!=\! 0.5$ pc.

For early-type galaxies we use the data of \cite{ben92} who list
$\sigma_{o}$ for a wide range of elliptical galaxies. In constructing this
data set, they take data from many different sources, who all provide
either mean-, global-, or central-velocity dispersions and who rarely, if
ever, state whether any aperture corrections were applied. So keeping
potential biases in mind we use their listed $\sigma$ values, which like
the dE/dE,Ns, are likely lower limits of the real $\sigma_{o}$ values.

\subsection{Colors, Metallicities and Photometric Mass-to-Light Ratios}
\label{ln:photoml}
To determine $\Upsilon_{V}^{\rm phot}$ a good knowledge of the ages and
metallicities of the stellar populations is required, for which basic
photometric and spectroscopic properties are needed. The apparent $V$-band
magnitudes listed for the clusters in Table~\ref{tab:gcobs} were taken
from \cite{hol99} (HCH99 designations) and \cite{pen04} (all other
clusters, see Table~\ref{tab:gcobs}) and were de-reddened based on the
quoted $E_{(B-V)}$ values taken from their respective papers.~For the
clusters from \cite{pen04}, the $E_{(B-V)}$ values ranged from 0.10 to
0.13 mag whereas for the HCH99 clusters a value of 0.11 mag was used based
on the reddening maps of \cite{bur82}. The de-reddened apparent $V$
magnitues ($V_{0}$) are listed in column 2 of Table~\ref{tbl:phot},
assuming $R_{V}$=3.1. No corrections for internal reddening in NGC
5128 were applied.

In Table~\ref{tbl:phot} we list the $(U\!-\!B)_{0}$, $(B\!-\!V)_{0}$, and
$(V\!-\!I)_{0}$ colours for the clusters from \cite{pen04}; however, for
all of the HCH99 clusters, only the $(V\!-\!I)_{0}$ colour was available.
The metallicities, $Z$, listed in column 7 of Table~\ref{tbl:phot} were
taken from \cite{bea08}, who performed a large spectroscopic study of the
globular cluster system of NGC~5128 in order to derive an empirical
metallicity distribution function calibrated on Milky Way GCs. The
metallicities were derived using a Tukey biweight scheme applied to six
Lick indices \citep[Fe4383, Mg$_{2}$, Mg\emph{b}, Fe5270, Fe5335, and
Fe5406; see][]{bea08}.~We use the de-reddened colours and metallicities to
compute $\Upsilon_{V}^{\rm phot}$ (see Table~\ref{tab:ds}) and compare
those to our dynamically derived mass-to-light ratios.

\begin{figure}
\plotone{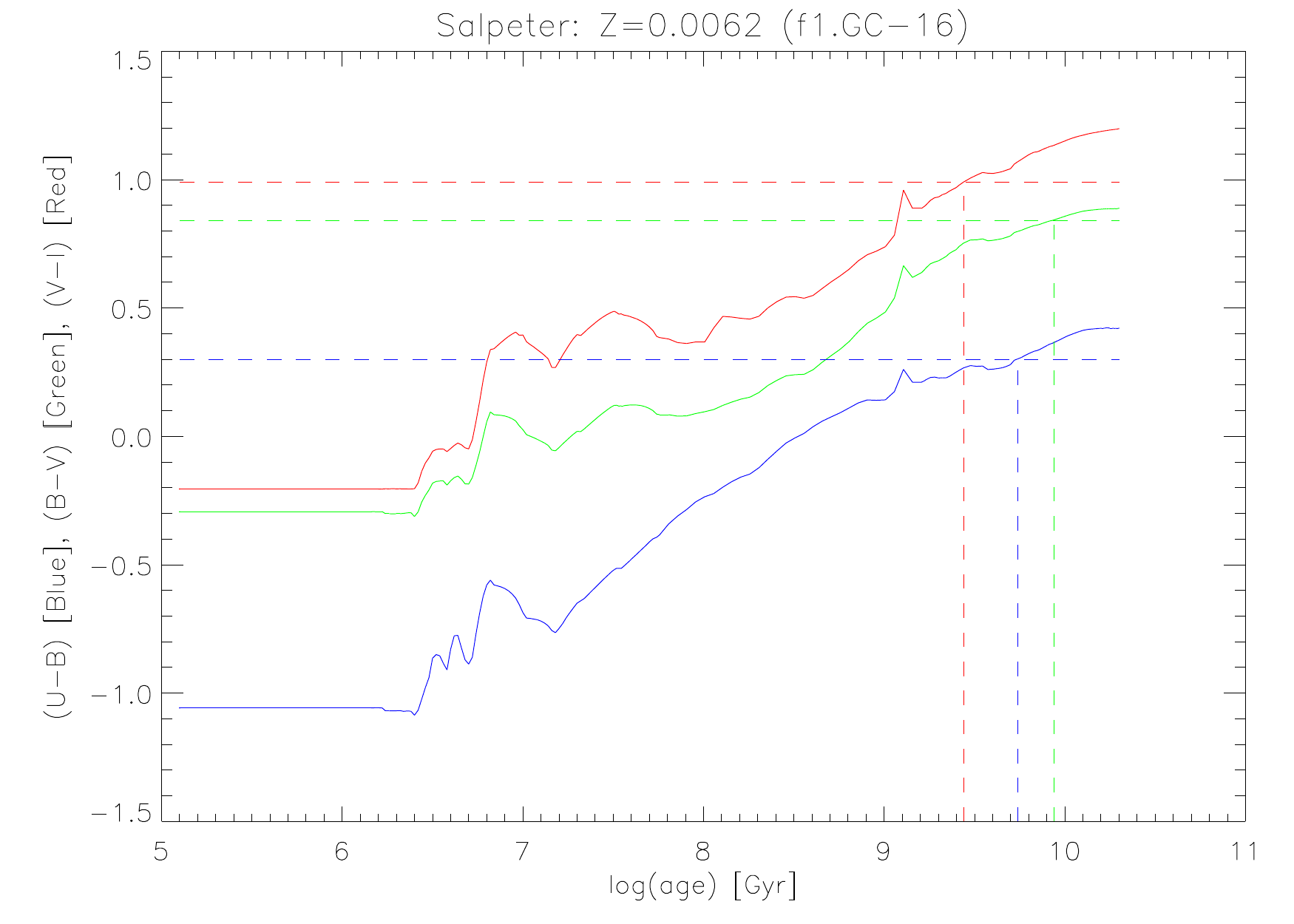}
\caption{An example plot showing the interpolated log(age) for one NGC~5128 GC (f1.GC-16/GC0378) from various colors. The derived mean age is then used in determining the photometric mass-to-light ratio $\Upsilon_{V}^{\rm phot}$. This technique has been applied to all GCs in our sample for which the photometric colors $U\!-\!B, B\!-\!V,$ and $V\!-\!I$ were available.}
\label{ps:photML}
\end{figure}

Owing to a lack of published age estimates based on spectroscopy in the
literature, we use spectroscopic metallicities which are listed in
\cite{bea08} for 16 star clusters from our sample to interpolate the SSP
models of \cite{bru03} to the corresponding metallicity, assuming a
\cite{cha03} IMF. We note that the metallicity scale of the Beasley et
al. measurements is tied to old Galactic GCs and for younger clusters
will systematically overestimate the metallicity values according to
$\Delta \log(Z)/\Delta \log(t)\simeq -1.0$ to $-2.0$ \citep{wor99}.
Figure~\ref{ps:photML} shows a plot of $(U\!-\!B)_{0}$, $(B\!-\!V)_{0}$,
and $(V\!-\!I)_{0}$ vs. $\log t$ for the cluster GC0378 (f1.GC-16) with
$Z\!=\!0.0062$ as an illustration of the method. The logarithmic ages
from the three colours $(U\!-\!B)_0$, $(B\!-\!V)_0$, and $(V\!-\!I)_0$
are $\log t\!=\!9.7$, 9.9, and 9.4, respectively, with the mean $\log
t\!=\!9.7\pm0.3$. This age translates to a photometric mass-to-light
ratio of $\Upsilon_{V}^{\rm phot}\!=\!2.2^{+0.1}_{-1.3}$, matching well
with our dynamically derived $\Upsilon_{V}^{\rm
dyn}\!=\!2.4^{+0.9}_{-0.7}$. This exercise was carried out for all 16
clusters and the output $\log t$ and $\Upsilon_{V}^{\rm phot}$ from all
interpolations are listed in Table~\ref{tbl:photML}. For comparison with
the \citeauthor{cha03} IMF results we also list the resulting $\log t$
and $\Upsilon_{V}^{\rm phot}$ assuming a \cite{sal55} IMF. From
Table~\ref{tbl:photML} it can be seen that $\Upsilon^{\rm phot}_{\rm
Sal}$ is on average 60\% larger than $\Upsilon^{\rm phot}_{\rm Chab}$.

The cluster ages based on $(V\!-\!I)_0$ alone were always found to be 
less than the mean of the ages when all three colours were used, and 
so because reliable values for $(U\!-\!B)_{0}$ and $(B\!-\!V)_{0}$ could 
not be found in the literature for the clusters GC0217 and GC0242, the 
photometric mass-to-light ratios for these two are considered to be highly 
uncertain.

\subsection{Dynamical Masses and Mass-to-Light Ratios}

While there are several ways of inferring the total mass of a compact
stellar system, the most straightforward approach is the virial theorem
\citep[e.g.][]{bt08}. This method relies on the symmetric properties of
what are assumed to be relaxed clusters with an isotropic stellar-orbit
distribution. The dynamical masses and mass-to-light ratios listed in
Table~\ref{tab:ds} are estimates based on the virial theorem of the form
\citep{spi87}:
\begin{equation}
\label{eq:virial}
{\cal M}_{\rm virial}\simeq2.5\,\frac{3\sigma^{2}_{o}r_{h}}{G}, 
\end{equation}
where $\sigma_{o}$ is the central line-of-sight velocity dispersion, and
3$\sigma^{2}_{o}$ is the mean-square velocity of the stars. $r_{h}$ is the
half-mass radius of the cluster measured in 3D, which if the assumption is
made that luminosity follows mass, can be inferred from the two
dimensional projected half-light radius,
R$_{h}$=$\frac{3}{4}r_{h}$.~Considering the treatment of the data sets
described in \S\S~\ref{ln:ap} and \ref{ln:ap_lit}, we therefore re-write
Equation~\ref{eq:virial} as
\begin{equation}
\label{eq:mdyn}
{\cal M}_{\rm dyn}=\beta \frac{\sigma^{2}_{o}R_{h}}{G}
\end{equation}
where the scaling factor $\beta$ depends on the surface brightness
profile.~We use this virial mass estimator for our subsequent dynamical
mass calculations. For galaxies with $R^{1/4}$ surface brightness profiles
\cite{cap06} find $\beta=5\pm0.1$, which corresponds to a S\'{e}rsic index
$n\approx5.5$ for an isotropic one-component system.~To stay consistent
with their study we use $\beta\!=\!5$ for our comparison sample of giant
early-type galaxies.~However, their measurements for M32 indicate that
$\beta$ is significantly higher for more compact stellar systems, in the
range $\beta\!\approx\!6\!-\!10$.~They also provide a scaling relation
between $\beta$ and the S\'{e}rsic index $n$ of the form
$\beta(n)=8.87-0.831n+0.0241n^2$ (cf. their Eq.~20) which implies
$\beta\approx7\!-\!8$ for our GCs and UCDs. A more detailed study of the
behavior of $\beta$ as a function of $n$ indicates a turnover at very low
$n$, typical for King profiles (C{\^o}t{\'e} \& McLaughlin, private
communication). For both our data and the UCDs with
$n\!\approx\!0.8\!-\!2.4$ we find $\beta\!\approx\!7.5$ to a very good
approximation.

With total mass estimates having been calculated, we estimate mass-to-light ratios from the de-reddened apparent $V$-band magnitudes
listed in Table~\ref{tbl:phot}.~The dynamical mass-to-light ratios
($\Upsilon_{V}^{\rm dyn}$) were computed by dividing the estimated virial
masses by the corresponding $V$-band luminosities
\begin{equation}
\label{eq:lum}
L_{V}=10^{-0.4[V-(m-M)_{V}-A_{V}-M_{V,\sun}]},
\end{equation}
assuming $M_{V,\sun}=4.83$ mag. The estimated $\Upsilon_{V}^{\rm dyn}$ 
for the 21 sample clusters, for which mass estimates could be made, range 
in value between 2.2 and 8.0 (see Table~\ref{tab:ds}).~The average
$\Upsilon_{V}^{\rm dyn}$ of the clusters is 4.2, significantly higher than 
the median $\Upsilon^{\rm dyn}_{V}$ of our M31 sample of $3.0\pm0.3$, 
and almost twice that of the median $\Upsilon^{\rm dyn}_{V}$ reported by 
\cite{mcl00} for Galactic GCs of $2.2\pm0.3$ using $\beta=10$ in 
Equation~\ref{eq:mdyn}, a value appropriate for the more compact LG 
GCs (see above) and thus assumed for this data set.

\section{Discussion}
\subsection{Comparison with Other Star Clusters}
\label{ln:comp}
\begin{figure}
\plotone{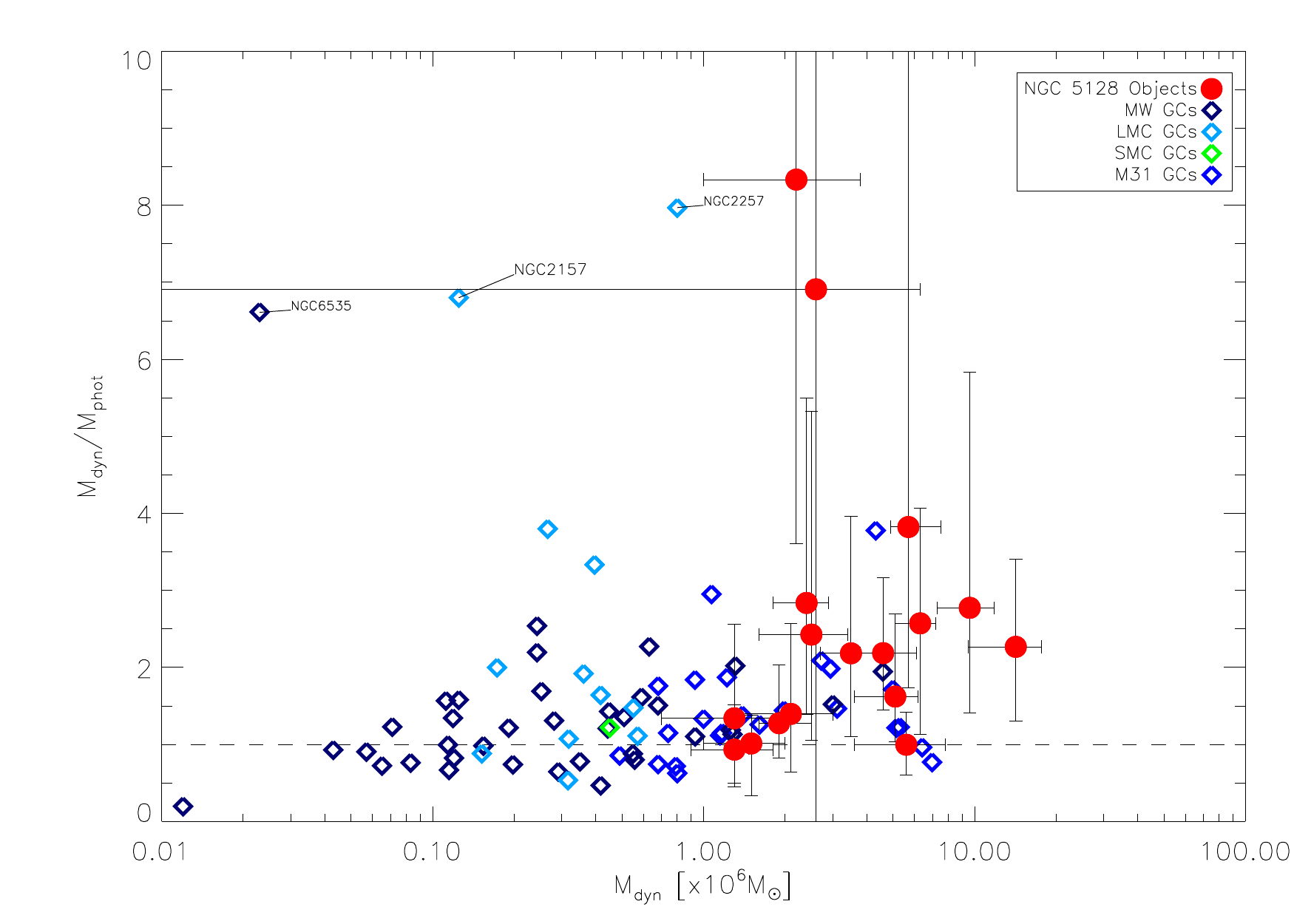}
\caption{${\cal M}_{\rm dyn}/{\cal M}_{\rm phot}$ vs. $\log({\cal M}_{\rm dyn})$ for NGC 5128 as well as Local Group GCs. The photometric mass-to-light ratios were computed from population synthesis model predictions as described in the text. Note the significantly increased dispersion in ${\cal M}_{\rm dyn}/{\cal M}_{\rm phot}$ for our dataset at ${\cal M}_{\rm dyn}\ga10^6\, M_\odot$ compared to the Local Group GCs.}
\label{fig:masscomp}
\end{figure}

To check for systematic differences between photometric and dynamical
mass-to-light ratios in our sample clusters, we plot in
Figure~\ref{fig:masscomp} the ratios of ${\cal M}_V^{\rm dyn}$ to ${\cal
M}_{V}^{\rm phot}$ against ${\cal M}_{\rm dyn}$ for our sample NGC~5128
objects and Local Group GCs. For the Local Group comparison sample
\cite{mcl05} derived structural parameters for SMC, LMC and the MW GCs using
three different profiles.~Values for $\Upsilon_{V}^{\rm dyn}$ based on the
\cite{wil75} profile were chosen because it fits cluster surface
brightness profiles at least as well, and often substantially better than
the \citeauthor{kin66} profiles or power-law models. M31 objects have ${\cal M}_{\rm dyn}$ derived from $\sigma_o$ listed in \cite{str09}, and $R_h$ from \cite{bar07} who also provide the $\Upsilon_{V}^{\rm phot}$ values used in 
Figure~\ref{fig:masscomp}.

Figure~\ref{fig:masscomp} indicates that while many of our sample clusters
have ${\cal M}_V^{\rm dyn}$ to ${\cal M}_{V}^{\rm phot}$ ratios consistent
with the bulk of Local Group GCs, our sample has on average ${\cal
M}_{\rm dyn}\simeq1.8\, {\cal M}_{\rm phot}$, or as much as
$\sim\!2.5-4.5$ for the cases of GC0265, GC0217 and GC0242. Possible
explanations for the particularly high ${\cal M}_V^{\rm dyn}$ to ${\cal
M}_{V}^{\rm phot}$ ratios include a dark gravitating component,
non-equilibrium dynamical processes (e.g.~tidal disruption,
pre-relaxation, young stellar systems), or/and simply that the
$\Upsilon_{V}^{\rm phot}$ values based on SSP models are underestimated
due to a younger-than-expected stellar sub-component or an unusual (top-
or bottom-heavy) IMF. The latter possibilty was proposed by \cite{dab08}
who show that the elevated $\Upsilon_{V}^{\rm dyn}$ of massive compact
objects with masses ${\cal M}_{\rm dyn}\ga2\times 10^6M_{\sun}$ cannot be
explained by current stellar population models if a canonical IMF is
assumed. They suggested that either a bottom-heavy IMF
\citep[see][]{kro08} or a top-heavy IMF \citep[see][]{dab09} could be
responsible for the discrepancy between dynamical and (canonical) stellar
population mass-to-light ratios. However for the cases of GC0265, GC0217
and GC0242, the high values shown in Fig.~\ref{fig:masscomp} are most
likely due to significantly underestimated photometric ages.

The possibility of a younger-than-expected stellar sub-component has some
support from recent work \citep[e.g.][]{bed04,pio08} that goes against the
current paradigm that GCs are always made of SSPs but show evidence for a
composite stellar population. However, if the ratio ${\cal M}_V^{\rm
dyn}/{\cal M}_{V}^{\rm phot}\approx2$ were entirely due to age and
metallicity this would translate to an age underestimate of $\Delta t\ga7$
Gyr {\it or} a metallicity underestimate of $\Delta$[Z/H]~$\ga 1$ dex.
Larger offsets would require a combination of even larger age {\it and}
metallicity mismatches. The spectroscopic metallicity estimates for our
sample GCs are expected to be accurate to better than $\sim\!0.3$ dex, but
our rough photometric ages (see Sect.~\ref{ln:photoml}) could be affected
by unknown horizontal-branch morphologies, significant blue-straggler
stellar populations, and variable chemical compositions, especially at
high metallicities and old ages.

We investigate the option of non-equilibrium stellar kinematics in the
form of signatures of tidal disruption in those clusters with high ${\cal
M}_V^{\rm dyn}/{\cal M}_{V}^{\rm phot}$ ratios, by studying the outliers
in the Local Group GC sample in Figure~\ref{fig:masscomp}. We find that
generally the clusters show indications of being in dynamically
non-equilibrium states. In the case of NGC 2157, the cluster is not
expected to be relaxed, having an age ($\log t = 7.60$ Gyr) less than its
median two-body relaxation time ($\log t_{\rm rh} = 8.85$ Gyr)
\citep{mcl05}. \cite{die02} performed a statistical study of LMC binary
clusters, and multiple cluster systems, finding that many of them could be
explained statistically by a chance superposition in the sky. However,
some tidally connected binary clusters may introduce non-virial velocity
fields which would lead to excess $\Upsilon_{V}^{\rm dyn}$. 

As a final check against the dependence of mass-to-light ratio
determinations on the choice of surface brightness profiles, the ${\cal
M}_{V}^{\rm dyn}/{\cal M}_{V}^{\rm phot}$ difference between
\citeauthor{wil75} and \citeauthor{kin66} profiles for the \cite{mcl05}
data was investigated. The results showed that although the
\citeauthor{kin66} profile results in systematically lower ${\cal
M}_{V}^{\rm dyn}/{\cal M}_{V}^{\rm phot}$ ratios, the median difference is
only $-0.08$ and does not affect our results significantly.

\subsection{Fundamental Plane Relations}
\label{ln:fp}
\subsubsection{$\Upsilon_{V}^{\rm}$ and ${\cal M}_{\rm dyn}$ Relations}
\label{ln:mlmass}
\begin{figure}
\epsscale{1.25}
\plotone{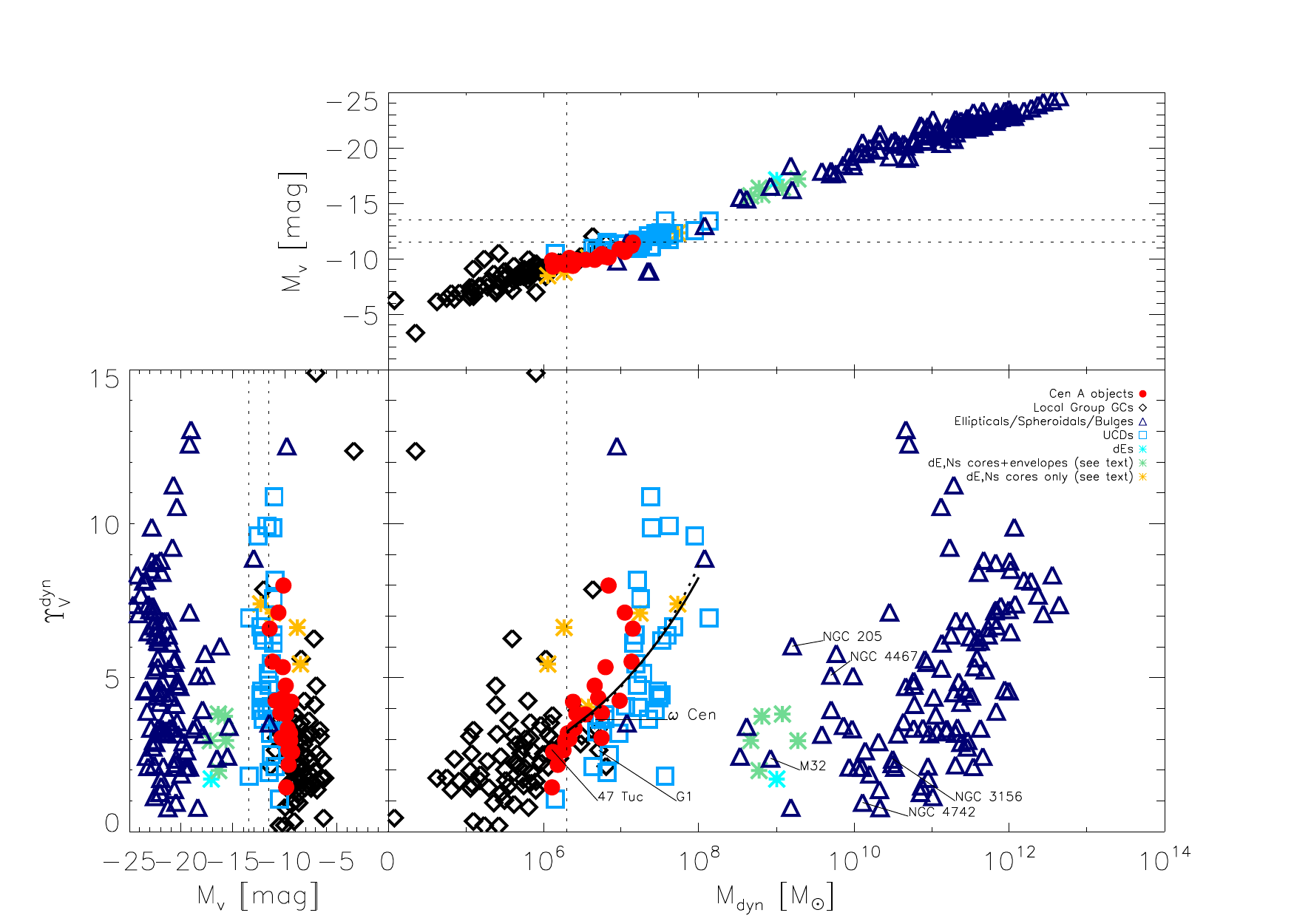}
\caption{Relations between the dynamical mass ${\cal M}_{\rm dyn}$, absolute
magnitude $M_{V}$ and the dynamical mass-to-light ratio
$\Upsilon_{V,\sun}^{\rm dyn}$ for compact stellar systems and early-type
galaxies. The Cen A objects ({\it solid circles}) have parameters derived
in this work. Local Group GC data ({\it open diamonds}) were taken from
\cite{mcl05} for LMC, SMC, MW clusters,
while the early-type galaxy data ({\it open triangles}) were culled from
\cite{ben92}. The dE/dE,N/core data ({\it asterisks}) were taken from
\cite{geh02}. The UCD data ({\it open squares}) are from three sources:
Virgo UCDs are taken from \cite{evs07}, and Fornax UCDs from \cite{mie08}
and Patrick C\^{o}t\'{e} (private communication). Several interesting
compact stellar systems are labeled. Dotted and solid curves are
exponential and power-law fits, respectively, as described in
Section~\ref{ln:fp}. Dotted lines indicate the mass break of \cite{has05}
at ${\cal M}_{\rm dyn}\simeq2\times10^{6}\, M_{\sun}$ and typical
luminosities for UCDs in the range $-13.5\!<\!M_V\!<\!-11.5$ mag from
\cite{hil09}.}
\label{fig:mdmvml}
\end{figure}

Figure~\ref{fig:mdmvml} shows the correlations between $\Upsilon_{V}^{\rm
dyn}$, ${\cal M}_{\rm dyn}$, and $M_V$, while Figure~\ref{fig:mvrhvd}
illustrates relations between $\sigma_o$, $R_h$, and $M_V$ for our sample
GCs as well as Local Group GCs, UCDs, and other early-type galaxies. In
Fig.~\ref{fig:mdmvml}, we see that clusters with $\Upsilon_{V}^{\rm
dyn}\ga5$ almost exclusively belong to objects with masses larger than
$\simeq\!2\times$10$^{6} M_{\sun}$, where most of our NGC~5128 GCs reside.
We find a strong correlation between $\Upsilon_{V}^{\rm dyn}$ and ${\cal
M}_{\rm dyn}$ and approximate the following two fitting functions:
\begin{eqnarray}
\log\Upsilon_{V}^{\rm dyn} &=& \Upsilon_{0,p}+
			\gamma\times \log{\cal M}_{\rm dyn} \label{fit:pow} \\
\log\Upsilon_{V}^{\rm dyn} &=& \Upsilon_{0,e}+
			\exp(\xi\times\log {\cal M}_{\rm dyn})  \label{fit:exp}
\end{eqnarray}
for compact stellar systems with $2\times10^{6}\!<\!{\cal M}_{\rm
dyn}\!<\!10^{8}\, M_{\sun}$ using Chauvenet criteria to remove
outliers.~For the power-law fit in Equation~\ref{fit:pow} we obtain
$\Upsilon_{0,p}\!=\!-0.99\pm0.15$ and $\gamma\!=\!0.24\pm0.02$ with a
reduced $\chi^2\!=\!2.10$.~For the exponential fit in
Equation~\ref{fit:exp} we get a marginally better reduced $\chi^2\!=\!2.08$ and find
$\Upsilon_{0,e}\!=\!-1.53\pm0.09$ and $\xi\!=\!(0.11\pm0.01)$. The
corresponding relations are illustrated in Figure~\ref{fig:mdmvml} as
dotted (exponential) and solid curves (power-law fit).~While the physical
cause for this correlation may be due to a non-baryonic component, we
point out that $\Upsilon_{V}^{\rm dyn}$ is also dependent on the cluster
SFH, the IMF, and any changes to the stellar population due to dynamical
evolution and the chemical makeup of the constituent stellar components.
The detailed discussion of the origin of this $\Upsilon_{V}^{\rm
dyn}\!-\!{\cal M}_{\rm dyn}$ trend must await more accurate spectroscopic
data.







\subsubsection{Size-Luminosity Relations}
\label{ln:mvrh}
\begin{figure}
\epsscale{1.25}
\plotone{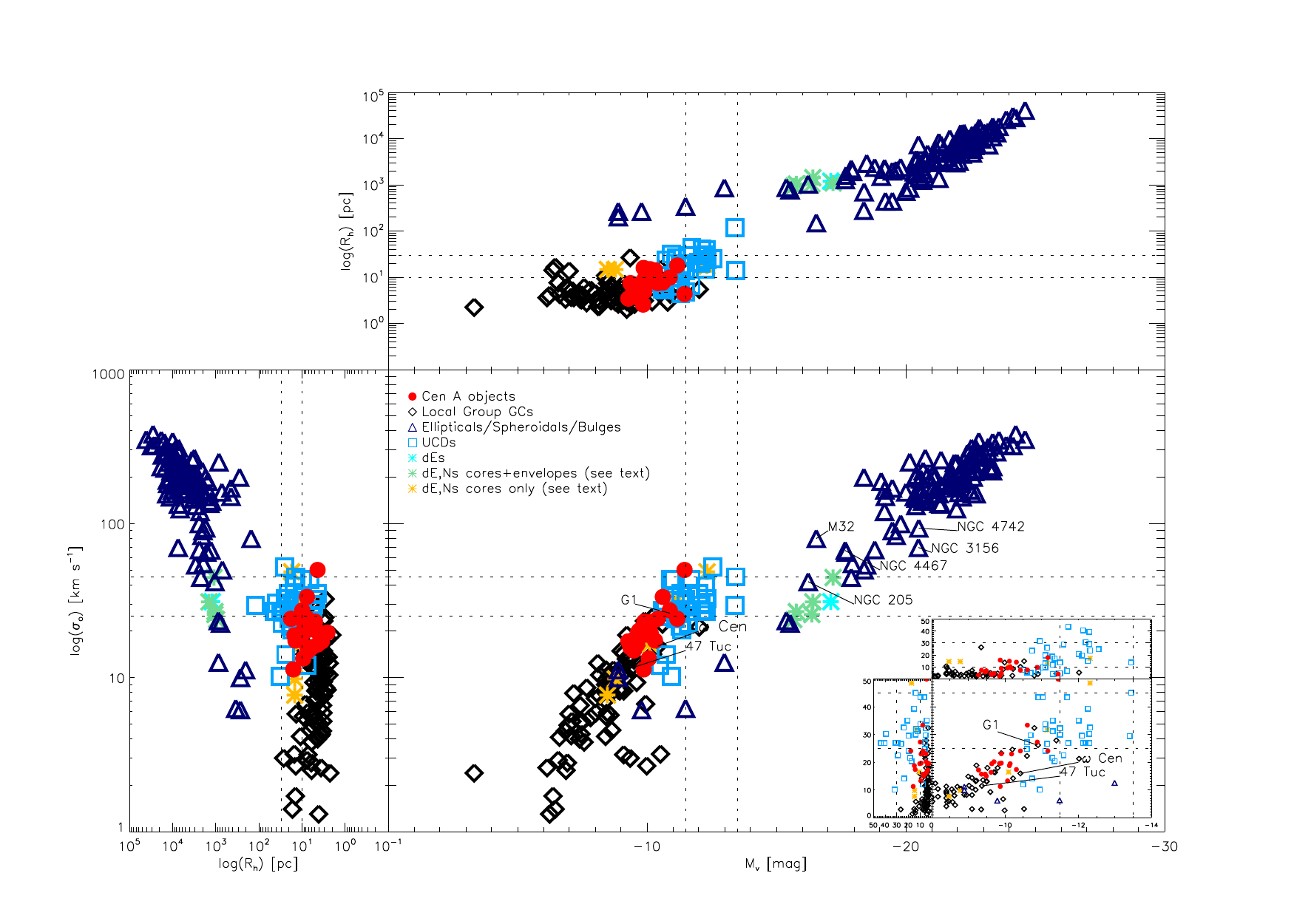}
\caption{Relations between absolute magnitude $M_{V}$, half-light radius $R_{h}$
and central velocity dispersion $\sigma_{o}$. The data samples in this
plot are identical to those given for Fig.~\ref{fig:mdmvml}. Several
interesting compact stellar systems are labeled. Dotted lines indicate the
typical luminosities, central velocity dispersions, and half-light radii
for UCDs in the range $-13.5\!<\!M_V\!<\!-11.5$ mag, $25<\sigma_o < 45$ km
s$^{-1}$, and $10<R_h<30$ pc from \cite{hil09}.}
\label{fig:mvrhvd}
\end{figure}

Fig.~\ref{fig:mvrhvd} clearly shows the break in the size-luminosity
relation in that the UCDs and dE,N cores all have sizes systematically
larger than the Local Group GCs for $-13.5\!<\!M_{V}\!<\!-11.0$ mag.
However, the NGC 5128 sample splits in the region
$-12.0\!\lesssim\!M_{V}\!\lesssim\!-10.0$ such that 4/12 (see inset of
Fig.~\ref{fig:mvrhvd}) objects appear to fall along the size-luminosity
relation for early-type galaxies, while the rest are more consistent with
no size-luminosity relation similar to Local Group GCs. We also note that
all of the NGC 5128 objects in this region show sizes that are more
consistent with those of low-luminosity UCDs and the dE,N cores, while
those objects having $M_{V}\!<\!-10.0$ show sizes similar to the Local
Group globular clusters. Figures~\ref{fig:mdmvml} and \ref{fig:mvrhvd}
also indicate that while the dE,N cores appear to be  similar to massive
GCs and UCDs, their envelopes are similar to low-luminosity early-type
galaxies \citep[see also][]{mho04,rej07}.

\cite{mie06} performed a spectroscopic study of compact stellar systems in
the Fornax Cluster to investigate possible formation scenarios for this
particular type of compact stellar object. A clear break in the
metallicity distribution was found at masses of $\sim\!3\!\times\!10^{6}\,
M_{\sun}$, and $M_{V}\!\approx\!-11$ mag such that for $M_{V}\!<\!-11$ mag,
the mean metallicity is $0.56\pm0.15$ dex higher than for $M_{V}\!>\!-11$
mag. A change in the size-luminosity relation for compact objects, in that
$R_{h}$ scales with luminosity for $M_{V}\!\lesssim\!-11$ mag and becomes
virtually independent of luminosity for $M_{V}\!\gtrsim\!-11$ mag, was
also reported to accompany the break in metallicity distribution,
suggesting the possibility of distinct formation scenarios \citep{mie06}.
\cite{mie08} recently revised this break-point to $M_{V}\!\approx\! -12$
mag and ${\cal M}_{\rm dyn}\simeq 2\times10^{6} M_{\sun}$, in agreement
with the previously reported mass break of \cite{has05}, and found that
systems with $\gtrsim2\times10^{6} M_{\sun}$ had $\Upsilon_{V}$ values roughly
twice as large as GCs of mass $\lesssim2\times 10^{6} M_{\sun}$. This is
in good agreement with our results.

\subsection{$\kappa$-space}
A common method of investigating the connection between globular clusters
and other spheroidal systems is through the adoption of $\kappa$-space
parameters. \cite{ben92} found that GCs and early-type
galaxies fall on two distinct planes in a three-dimensional space defined
by the parameters:
\begin{equation}
\kappa_{1}\equiv\frac{\log(\sigma_{o}^{2})+\log(R_{h})}{\sqrt{2}}
\end{equation}
\begin{equation}
\kappa_{2}\equiv\frac{\log(\sigma_{o}^{2})+2\log(I_{e})-\log(R_{h})} 
{\sqrt{6}}
\end{equation}
\begin{equation}
\kappa_{3}\equiv\frac{\log(\sigma_{o}^{2})-\log(I_{e})-\log(R_{h})} 
{\sqrt{3}}
\end{equation}
where $I_{e}\!=\!L_{V}/(2\pi R_{h}^{2})$ and $R_{h}$ is the projected
half-light radius.~As can be seen above, the parameters $\kappa_{2,3}$ are
defined by luminosity surface density. However, to compare GCs to younger
clusters and galaxies, it is more useful to define $\kappa$-space using
mass surface density \citep[see][]{mcl07} and to convert $R_{h}$ to kpc
\citep[see][]{ben92}. To compute this we use
$\log\Sigma_{h}=\log\Upsilon_{V}+\log I_{e}$, so that the new
$\kappa$-space is defined by:
\begin{equation}
\hat{\kappa}_{1}=\frac{\log(\sigma_{o}^{2})+\log(\frac{R_{h}}{1000})}{\sqrt{2}}
\end{equation}
\begin{equation}
\hat{\kappa}_{2}=\frac{\log(\sigma_{o}^{2})+2\log(\Sigma_{h})-\log(\frac{R_{h}}{1000})}{\sqrt{6}}
\end{equation}
\begin{equation}
\hat{\kappa}_{3}=\frac{\log(\sigma_{o}^{2})-\log(\Sigma_{h})-\log(\frac{R_{h}}{1000})}{\sqrt{3}}
\end{equation}

\begin{figure}
\epsscale{1.25}
\plotone{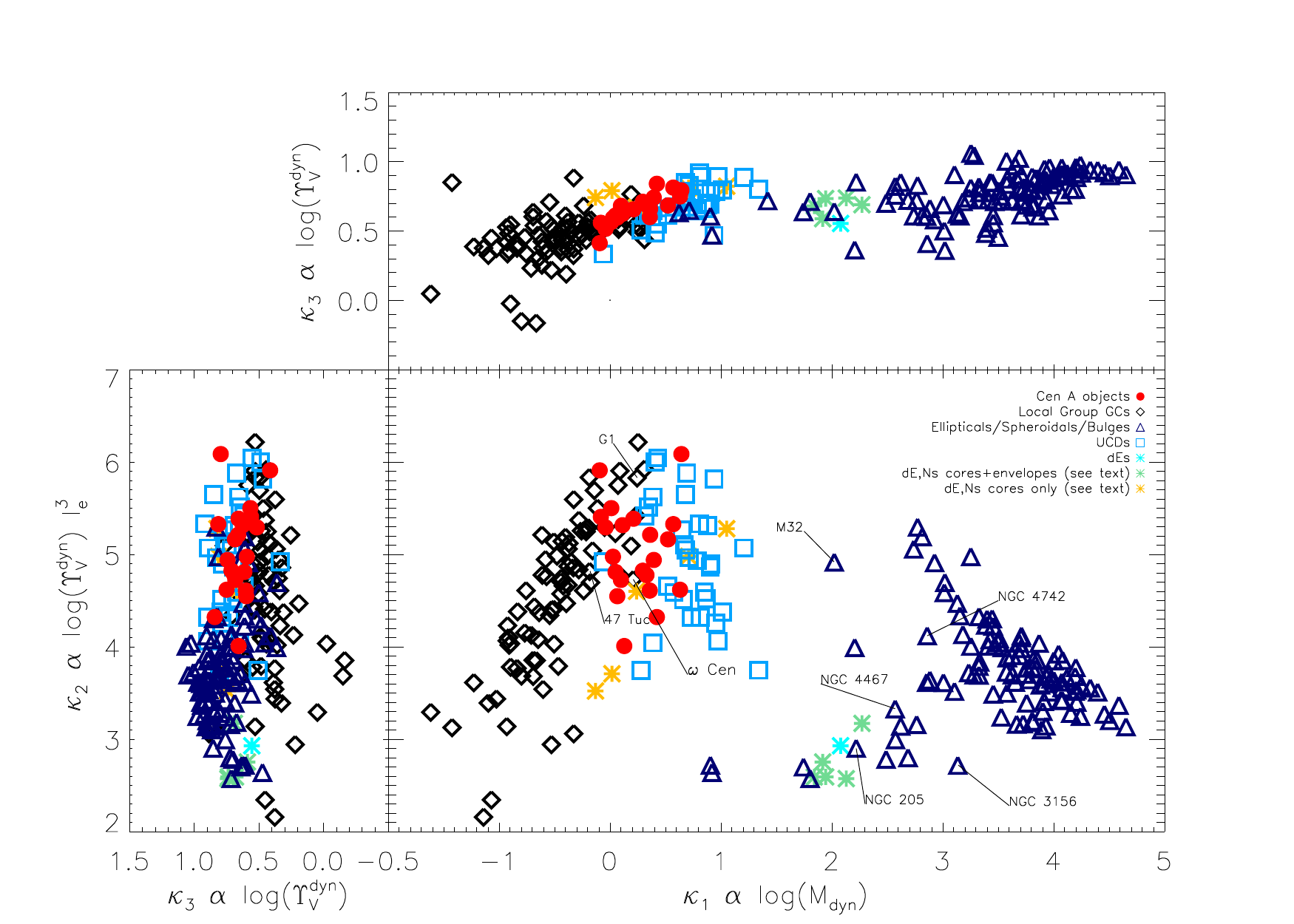}
\caption{$\hat\kappa$-space for our Cen A sample GCs, Local Group GCs, UCDs and
early-type galaxies. The sources in this plot are identical to those given
for Fig.~\ref{fig:mdmvml}. Several interesting compact stellar systems are
labeled.}
\label{fig:kap}
\end{figure}

Figure~\ref{fig:kap} shows the $\hat{\kappa}$-space with the above mass
surface density formalism. The data used for this plot are identical to those used
for Figs.~\ref{fig:mdmvml} and \ref{fig:mvrhvd} (see \S\S~3.3-3.5). In
order to ensure that all of the combined data are as consistent as
possible, only the most central measurements were taken from the
respective papers. It was found that by simply culling $\sigma_{o}$,
$R_{h}$ and absolute $V$ magnitude, $M_{V}$ from the sources enabled
$\hat{\kappa}_{1,2,3}$ to be calculated in a consistent manner. Using
Equation~\ref{eq:mdyn} to calculate ${\cal M}_{\rm dyn}$ combined with
$M_{V}$ gives $\Upsilon_{V}^{\rm dyn}$, which with $I_{e}$ provides
$\Sigma_{h}$, yielding all of the parameters necessary to construct the
$\hat{\kappa}$-space. The fundamental plane for GCs is typically analyzed
using population synthesis models, which provide $\Upsilon_{V}^{\rm phot}$
that is combined with other structural parameters to construct the
fundamental plane. However, in the conversion from luminosity to mass
surface density for $\hat{\kappa}_{2,3}$, we use $\Upsilon_{V}^{\rm dyn}$
rather than $\Upsilon_{V}^{\rm phot}$ for the reasons discussed in
\S\ref{ln:comp}.

In the $\hat{\kappa}_{1,2}$ space, there is a clear separation between the
Local Group GCs and the elliptical galaxies \citep[see also][]{mho04}.
Figure~\ref{fig:kap} shows the two distinct planes corresponding to
`classical' GCs and galaxies ranging in mass from dE and dSph to giant
ellipticals. Our NGC 5128 clusters appear systematically more massive than
the Local Group GCs, with many encroaching upon the region of
$\hat{\kappa}$-space inhabited by the UCDs, and even sharing the space
containing the nuclei of dE,Ns studied by \cite{geh03}. The UCDs, being
typically more massive than the NGC 5128 objects, fall clearly within the
gap, suggesting a common formation mechanism between the most massive GCs
and other compact stellar systems. The possibility that some massive GCs
really are the remnants of dE,Ns that have been tidally stripped may have
some support from the discrepancy between $\Upsilon_{V}^{\rm dyn}$ and
$\Upsilon_{V}^{\rm phot}$, as some of these objects may be in the process
of having their dark matter halos stripped.

\subsection{Stellar Population Properties}
In this section we investigate the chemical compositions of our sample of
NGC 5128 systems by comparing absorption line strengths of several age and
metallicity indicators to those of `classical' GCs, UCDs and various
early-type galaxies. The objects analyzed in this section are a sub-sample
of those used in the above dynamical analysis and include all of the same
objects for which both chemical enrichment and dynamical data are 
available. We point out that we compare here dynamical properties with
the chemical makeup of our target clusters that were obtained with two
different instruments. It is, therefore, likely that these measurements 
sample different physical regions. Hence, our subsequent analysis may be
prone to dynamical and stellar-population gradients in the clusters.
However, for most Local Group GCs this will not affect the results as
dynamical relaxation will wipe out any radial gradient within a few
Gyr.~Nevertheless, we keep in mind that for more massive and extended
systems \citep[$t_{\rm rh}\!\propto\! {\cal M}^{1/2}R_h^{3/2}$, see
e.g.][]{bt08} this may introduce biases if radial stellar-population
gradients exist.

\subsubsection{The Absorption Line Index Data}
In order to investigate the chemical compositions, we have chosen to adopt
the Lick/IDS standard system.~The Lick system was originally introduced by
\cite{bur84} to investigate elemental abundances when using
intermediate-resolution integrated light spectra of massive,
extra-galactic stellar systems. 

The Lick data used for this analysis come from multiple sources and are as
follows. Data for the NGC 5128 clusters were taken from \cite{bea08} who
provide measurements of 23 indices for a large sample of the NGC 5128 GC
population all of which have had the relevant corrections applied and have
been calibrated to the passband definitions of \cite{wor94} and
\cite{wor97}.

Data for a total of 39 LG GCs were gathered from many different sources.
Measurements of 16 indices were taken from \cite{puz02}, \cite{bea02},
and \cite{sch05} for the 23 MW GCs, while values for 20 indices are
provided by \cite{bea02} for six GCs associated with the LMC. Ten
additional GCs, eight of which are associated with M31 were added to the
LG sample from \cite{bea04} and \cite{tra98}; however, it should be
noted that the values for the four indices H$\delta_{A,F}$ and
CN$_{1,2}$ were not always available for these ten clusters. In all
cases except for data from \cite{sch05}, the data were fully corrected
and calibrated to the passband system of \cite{wor94} and \cite{wor97}.
Due to the complicated nature of measuring the Lick indices of
extra-galactic objects, it is not always possible to obtain reliable
measurements of all 25 line indices for every object in any given study
as evidenced by the combined data set described above. In the case of
the \cite{sch05} data, measurements were available for only 16 indices;
however \cite{puz02} and \cite{bea02} provide some of the missing
measurements for five overlapping clusters, and so those values are used
where necessary.

With the Local Group GCs populating the lower end of the mass range for a
comparative sample, Lick data for early-type galaxies were taken from
\cite{tra98}, \cite{oga08} and \cite{kun00}. A total of 76 objects from
the \cite{tra98} dataset was found to overlap with objects for which we
have dynamical data derived in this study, and another eight objects were
provided by \cite{oga08}. \cite{tra98} provide measurements for 19 indices
with seven objects missing H, Fe, and Mg index measurements in varying
combinations. The data provided by \cite{oga08} include 11 index
measurements, with only the H$\beta$, three Mg, six Fe, and the NaD
indices available. Neither \cite{tra98} nor \cite{oga08} provide
measurements of the H$\gamma_{A}$ and H$\gamma_{F}$ indices, although it
was found that they were measured by \cite{kun00} for four of the
\cite{tra98} galaxies. We therefore include those data in any relevant
plots.

Analogous to the previous dynamical comparisons, data for objects that
bridge the mass-gap between GCs and early-type galaxies were culled from
\cite{evs07} and \cite{geh03}. Specifically, measurements of the
H$\beta$, Mgb, Fe5270, Fe5335 indices as well as the [MgFe]$'$ and
$<$Fe$>$ indicators for seven UCDs and five dE,Ns in the Virgo cluster
are taken from those two papers respectively. For both sets of objects,
the line widths were calculated using the passband definitions of
Worthey 94/97 with the spectra degraded to the Lick resolution and all
corrections applied according to the prescription given in that paper.
The data for the dE,Ns were not corrected for internal velocity
dispersions; however the corrections themselves were significantly
smaller than the broadening function used \citep{geh03}.

\subsubsection{Chemical Composition of Stellar Systems}

\begin{figure}
\epsscale{2.0}
\plottwo{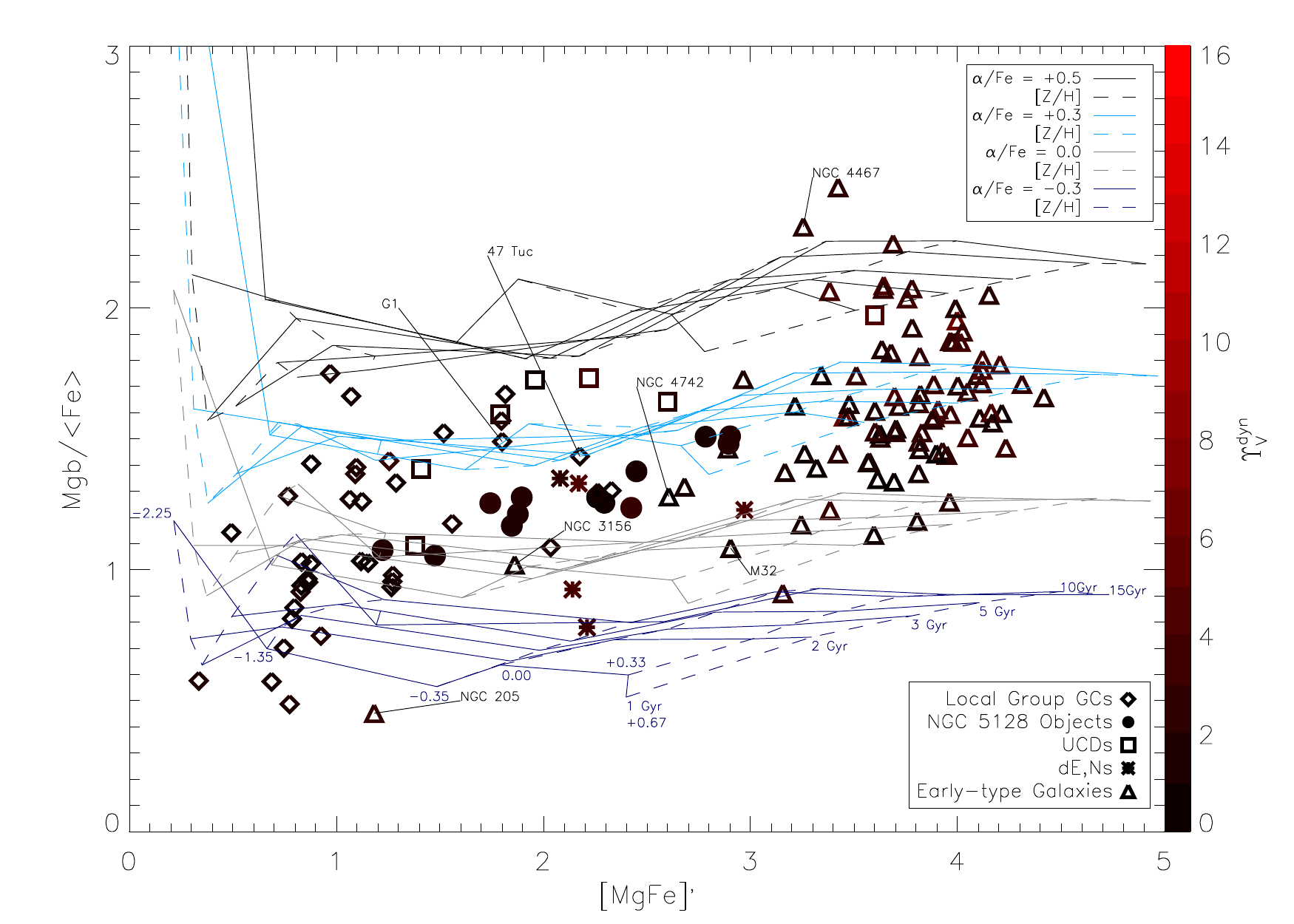}{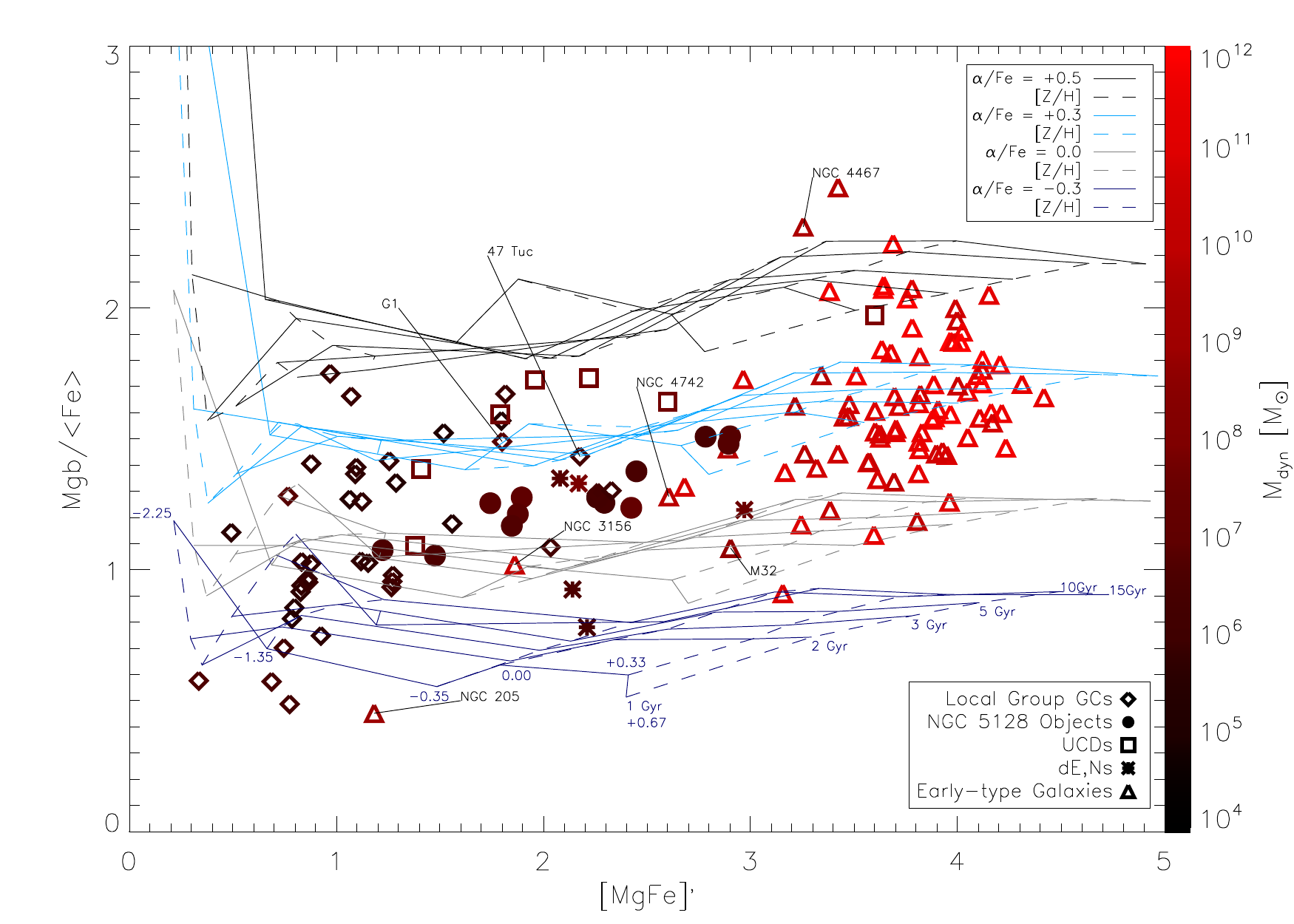}
\caption{[$\alpha$/Fe] diagnostic plot for NGC~5128 GCs and other compact stellar
systems. Individual datasets are indicated in the legend (lower right
corner) and described in the text in more detail. The symbol colors
parameterizes the  $\Upsilon_{V}^{\rm dyn}$ (top panel) and ${\cal M}_{dyn}$
(bottom panel).~We use the population synthesis model predictions from
\cite{tho03} for a range of ages ({\it solid lines}: $1\!-\!15$ Gyr),
metallicities ({\it dashed lines}: [Z/H]~$=\!-2.25$ to $+0.67$ dex), and
[$\alpha$/Fe] ratios ({\it grid colors}: $-0.3$ to $+0.5$ dex, see
legend). Note that each grid is almost entirely degenerate in age and
metallicity which enables us to determine accurate [$\alpha$/Fe] ratios
without the need of very accurate ages and metallicities. Several
interesting compact stellar systems are labeled.}
\label{fig:mgbfemgfe}
\end{figure}

\begin{figure}
\epsscale{2.0}
\plottwo{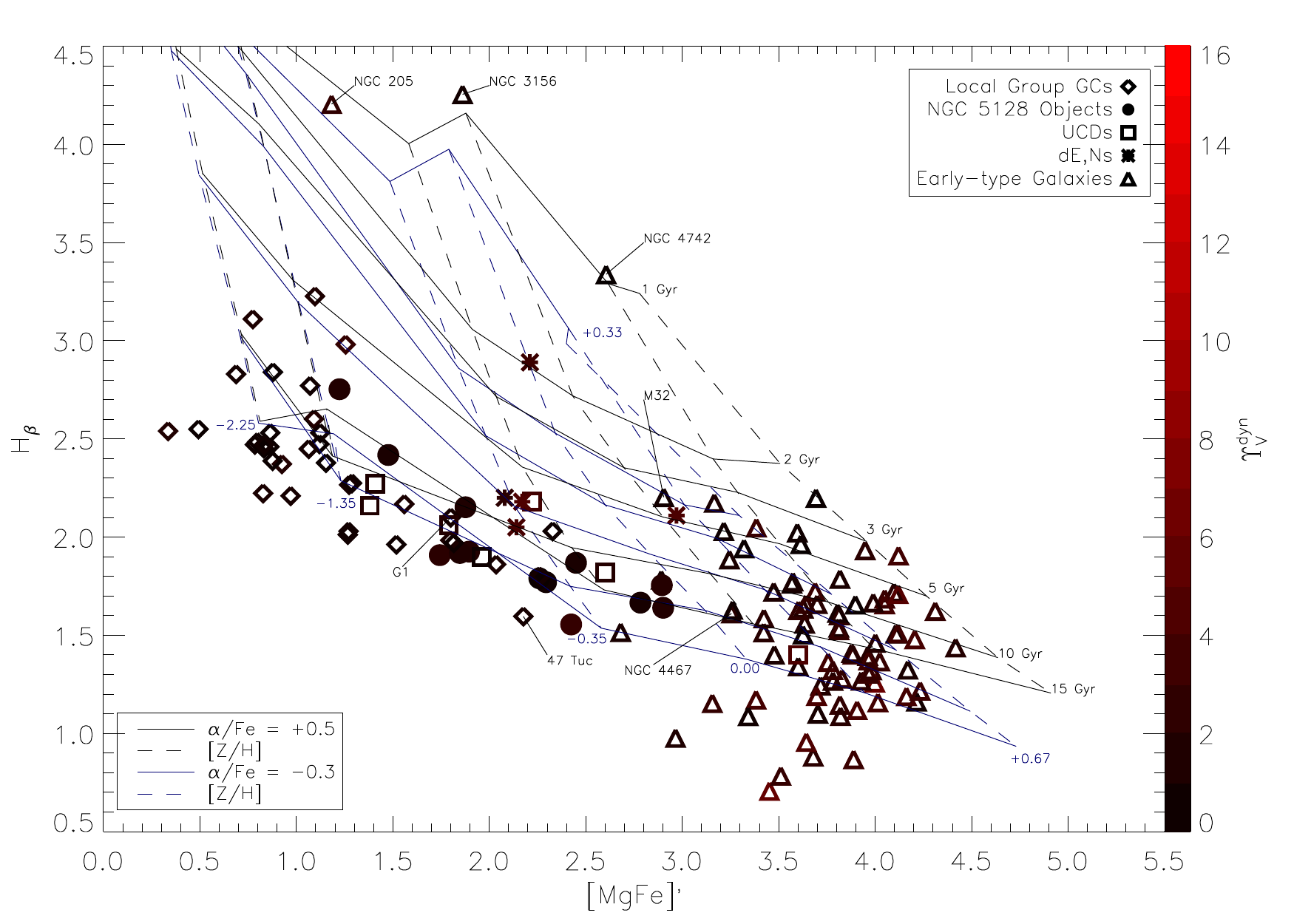}{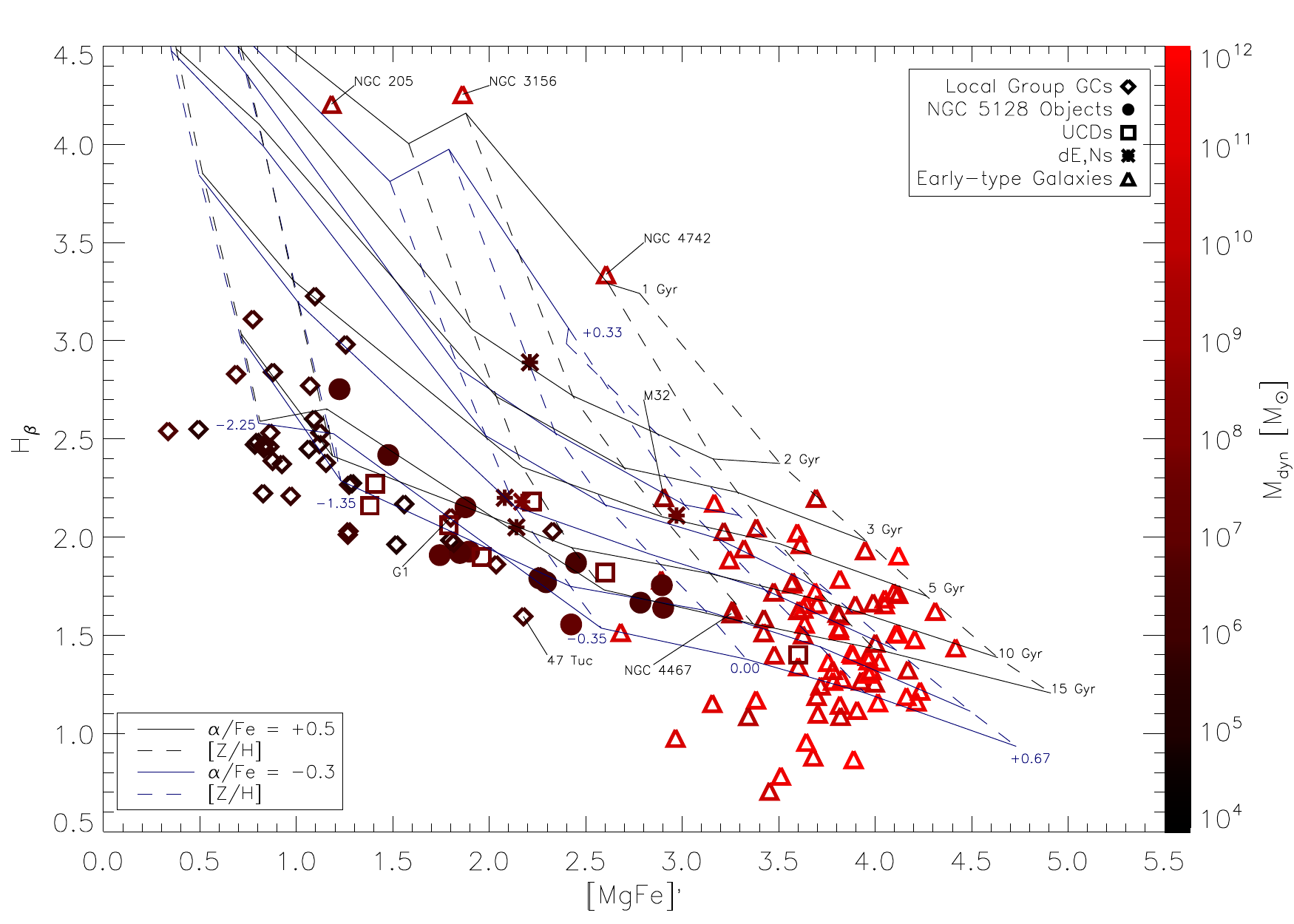}
\caption{Age-metallicity diagnostic diagram for the same sample as in
Figure~\ref{fig:mgbfemgfe}. Population synthesis model predictions are
from \cite{tho03} for a range of ages ({\it solid lines}: $1\!-\!15$ Gyr),
metallicities ({\it dashed lines}: [Z/H]~$=\!-2.25$ to $+0.67$ dex), and
two [$\alpha$/Fe] ratios ({\it grid colors}: $-0.3$ and $+0.5$ dex, see
legend in lower left corner). Symbol colors are parameterized by the
corresponding $\Upsilon_{V}^{\rm dyn}$ (top panel) and ${\cal M}_{dyn}$ (bottom
panel). Several interesting compact stellar systems are labeled.}
\label{fig:hbmgfe}
\end{figure}

\begin{figure}
\epsscale{2.0}
\plottwo{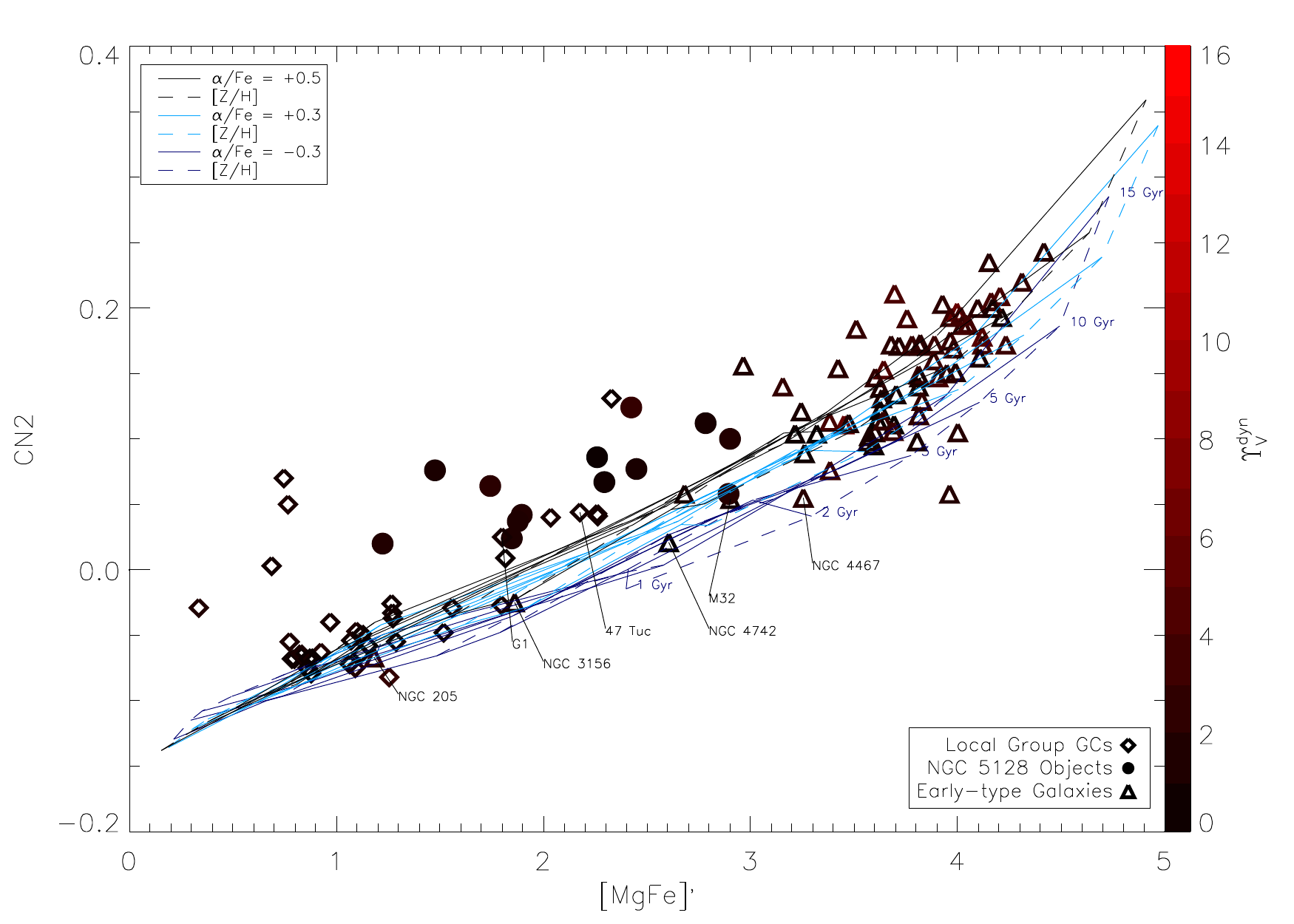}{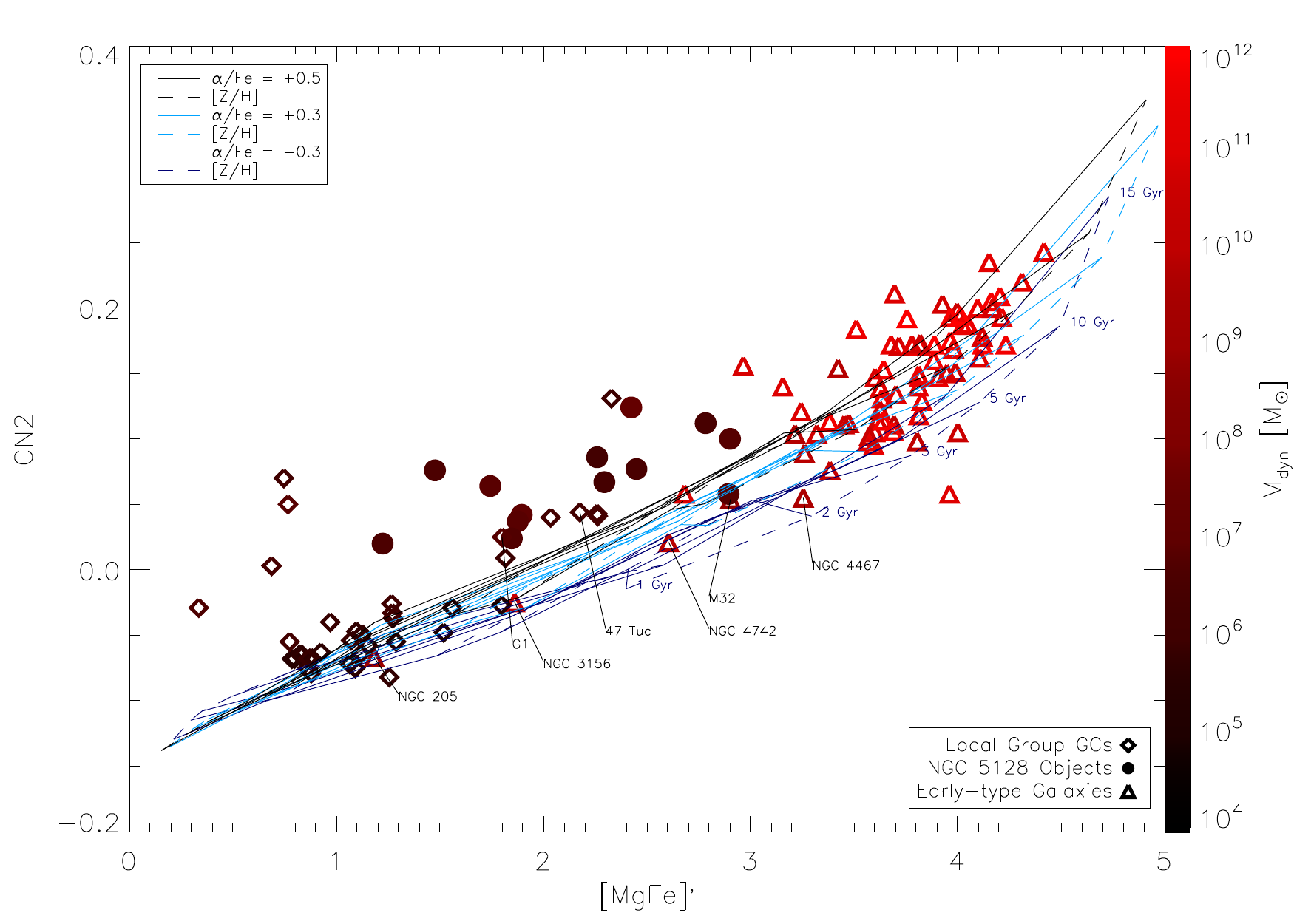}
\caption{CN$_2$ vs. [MgFe]\arcmin\ diagnostic plot. Model grid, symbol colors, and
dataset are as in Figure~\ref{fig:hbmgfe}. However, note that due to the
limited wavelength coverage of the UCD spectra in the blue we miss the
CN$_2$ for those objects. Note the full degenerate state of the model grid
in age, metallicity, and [$\alpha$/Fe]. This facilitates, at least
qualitatively the determination of a CN over- or under-abundance.}
\label{fig:cn2mgfe}
\end{figure}

\begin{figure}
\epsscale{2.0}
\plottwo{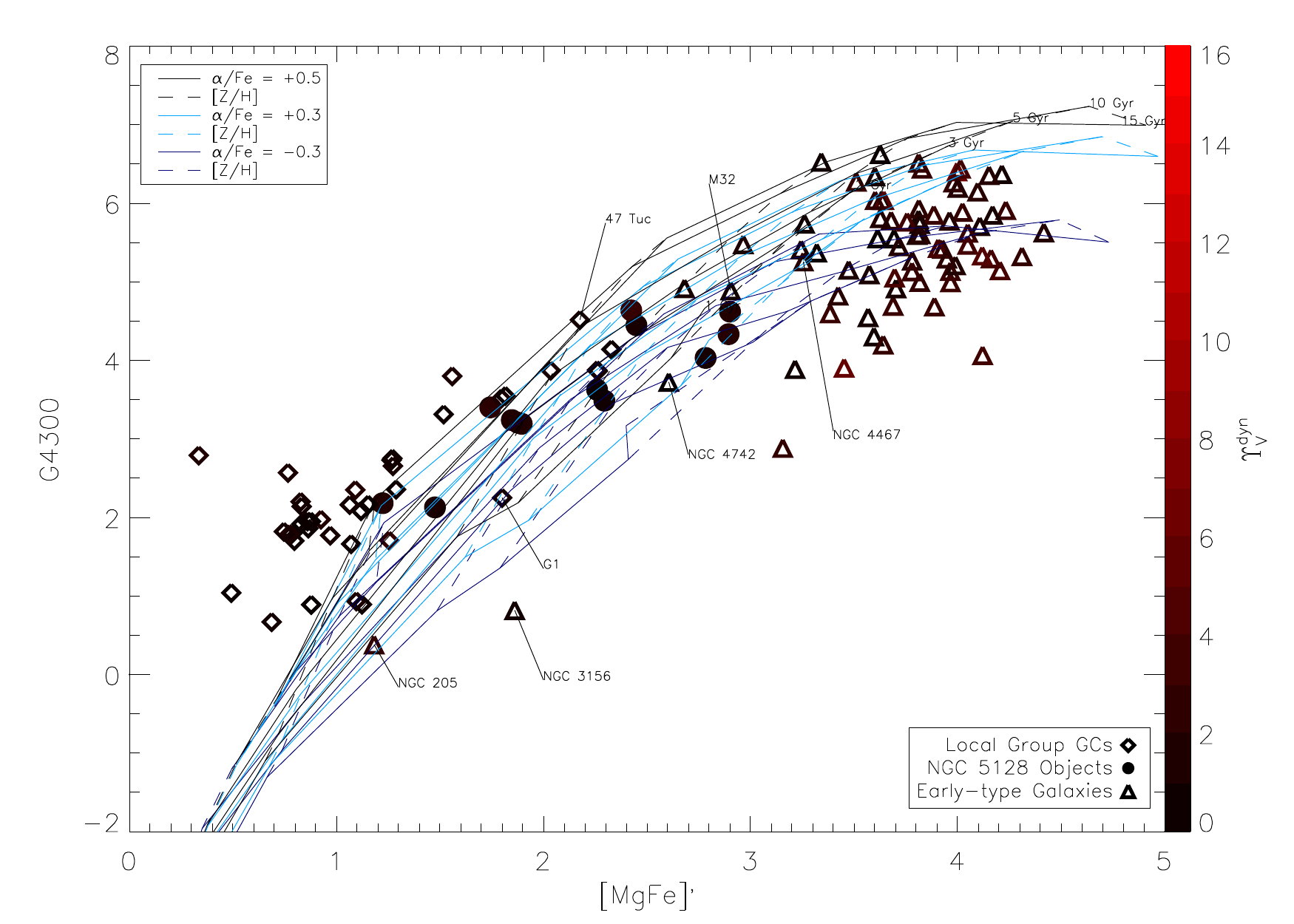}{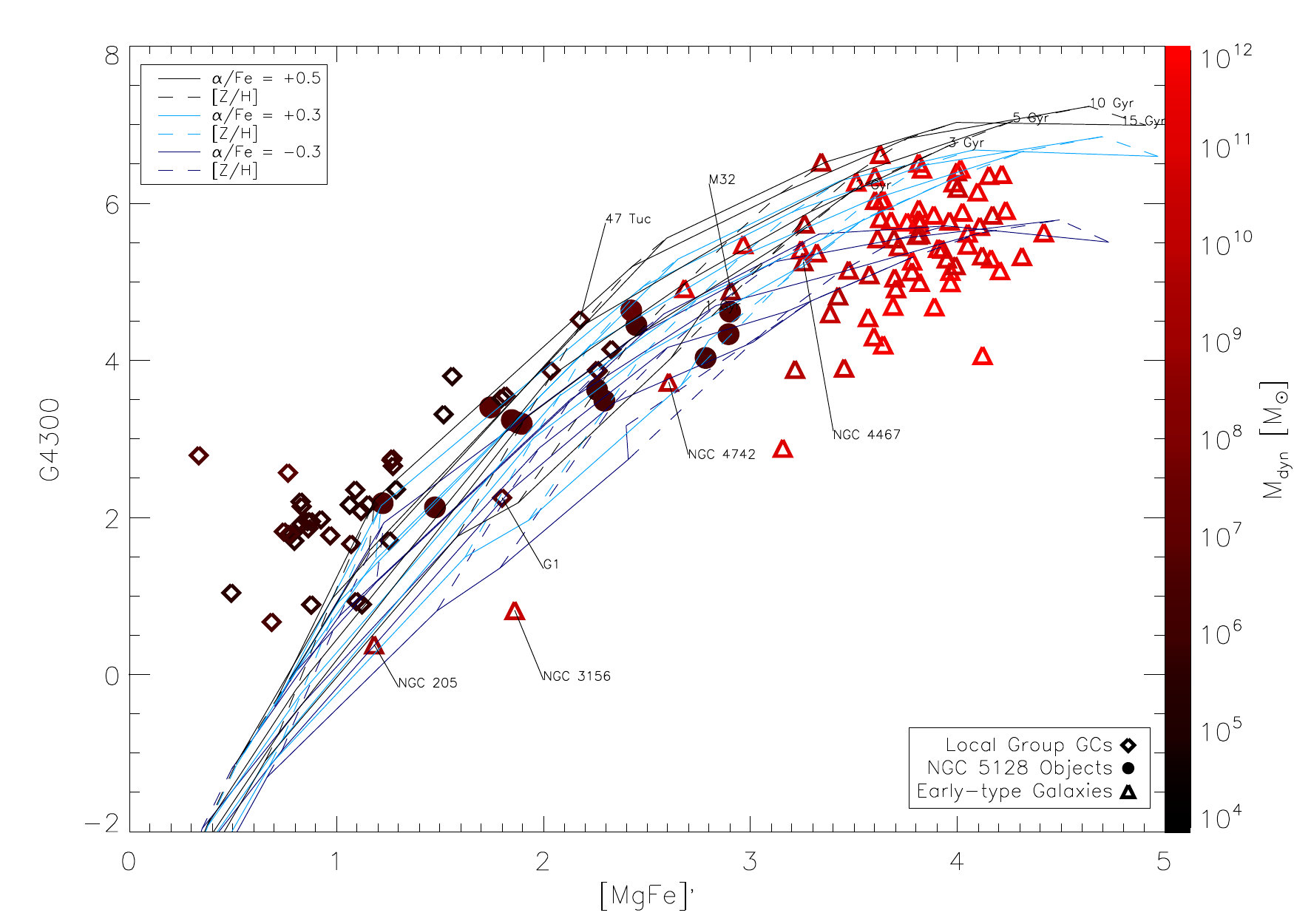}
\caption{G4300 vs. [MgFe]\arcmin\ diagnostic plots. G4300 index is predominantly tracing
the carbon abundance. Model grid, symbol colors, and dataset are as in Figure~\ref{fig:cn2mgfe}. The grid is degenerate in age, metallicity, and
[$\alpha$/Fe]. At [MgFe]\arcmin$\approx1.5\!-\!3.0$ \AA\ the model
predictions match the data and combined with abundance of CN$_{2}$ (see Fig.~\ref{fig:cn2mgfe}) indicate a nitrogen enhancement.}
\label{fig:g4300mgfe}
\end{figure}

Figures~\ref{fig:mgbfemgfe}-\ref{fig:g4300mgfe} show various
diagnostic diagrams for the sample described above, with the colour
gradients parameterized by $\Upsilon_{V}^{\rm dyn}$ (top panels) and
${\cal M}_{\rm dyn}$ (bottom panels). In all cases, the plots are shown with
the SSP model tracks of \cite{tho03} for different ages, [Z/H] and
[$\alpha$/Fe] ratios.~SSP isochrones are shown for ages of 1, 2, 3, 5, 10
and 15 Gyrs and for [$\alpha$/Fe] values of $-0.3, 0.0, 0.3,$ and 0.5 dex.
For each isochrone and iso-[$\alpha$/Fe] track, dashed lines of
iso-metallicity are shown for [Z/H]~=~$-2.25, -1.35, -0.35,$ 0.00, 0.33,
and 0.67 dex.

Balmer lines are often used as age indicators in this kind of diagnostic
because of the assistance they provide in breaking the age-metallicity
degeneracy of SSP models \citep[e.g.][]{puz03, sch05}. The Lick system
provides definitions for five Balmer lines (H$\beta$, H$\delta_{A}$,
H$\delta_{F}$, H$\gamma_{A}$, H$\gamma_{F}$), out of which we use H$\beta$
as the best age indicator due to both the sparse availability of H$\delta$
and H$\gamma$ measurements for early-type galaxies and the relative
insensitivity to [$\alpha$/Fe] that it shows compared to the other Balmer
lines \citep{tho04}.

The star formation histories of stellar populations are often inferred
through [$\alpha$/Fe] estimates. $\alpha$-elements are produced
predominantly by type-II supernovae early in the star formation history of
the population, while Fe-peak elements have a more drawn out enrichment
history due to the longer lifetimes of type-Ia SN progenitors
\citep[e.g.][]{gre08}. Thus, super-solar abundance ratios should be
expected of populations that have experienced a brief history of star
formation and are typical of early-type galaxies \citep{tra00, tho05},
while solar and sub-solar abundance ratios indicate more prolonged
chemical enrichment processes.

Figure~\ref{fig:mgbfemgfe} shows a plot of the index ratio
Mg$b/\langle$Fe$\rangle$, where $\langle$Fe$\rangle\!=\!({\rm Fe}5270+{\rm
Fe}5335)/2$, against the [$\alpha$/Fe]-insensitive metallicity indicator
[MgFe]\arcmin~$\!=\! \{ {\rm Mg}b\cdot(0.72\;{\rm Fe}5270+0.28\;{\rm
Fe}5335)\}^{0.5}$ \citep{tho03}. It can be seen that the UCDs (open
squares) almost exclusively have super-solar [$\alpha$/Fe] ratios between
+0.3 and +0.5 dex, with only one indicating a solar abundance ratio, while
the majority of the NGC 5128 objects (filled circles) and dE,Ns
(asterisks) indicate significantly less $\alpha$-enhanced stellar
populations with some of the dE,Ns even indicating sub-solar ratios. The
discrepancy between these populations is most likely due to different star
formation histories. However, it has also been suggested that a
combination of changing IMF slope and binary fraction could have a
significant effect on [$\alpha$/Fe] ratios \citep[e.g.][]{tra98, puz03}.

In Figure~\ref{fig:hbmgfe} we show the age-sensitive Balmer index
H$\beta$ plotted against [MgFe]\arcmin. Local Group GCs, UCDs and NGC
5128 objects all appear to share similarly old ages ($\ga$ 8 Gyrs).~More
interesting are the metallicities of the UCDs and NGC 5128 objects which
are significantly lower than those of the ellipticals, yet similar to
the most metal-rich Local Group GCs. If these objects represent the
stripped remains of dE,Ns, then they should show metallicities similar
to their progenitor dE galaxies, unless the objects were stripped of
their gas before their initial star formation was complete
\citep[i.e.][]{mie06}. In this case, there would be no opportunity for
self-enrichment, and their metallicities would tend toward lower values
in the present day unless the protocluster environments were such that
the objects were pre-enriched to higher metallicities before the onset
of star formation processes that created the star cluster \citep{bai09}.

Figures~\ref{fig:cn2mgfe} and \ref{fig:g4300mgfe} show plots of CN$_{2}$
and G4300 as functions of [MgFe]\arcmin. There are two CN indices,
CN$_{1}$ and CN$_{2}$, defined by the Lick system and together they trace
the strength of the CN absorption band at 4150\AA\ \citep{puz03}.~While
both CN indices give similar results, CN$_2$ avoids the H$\delta$ line
which leads to a more accurate prediction of the CN index strength,
especially for 1~$\lesssim$~Mg$b~\lesssim$~2~\AA\ \citep{tho03}. For this
reason we choose to use CN$_{2}$ for the diagnostic plots. The G4300 index
mainly traces the abundance of carbon in the G band, but also serves to
trace oxygen at $\sim1/3$ of the carbon sensitivity for giant ellipticals.
When shown together with the model tracks in Figure~\ref{fig:cn2mgfe}, it
can be seen that the NGC~5128 objects appear to have CN abundances
consistent with, if not slightly higher than, the Local Group GCs. On
average they lie above the model tracks when compared to early-type
galaxies at the same metallicities, which can be interpreted as an
over-abundance of CN. It is interesting to note that a large population of
CN-enriched stars has been observed in $\omega$Cen \citep[][for
example]{hil00}, hinting that it may be related to other, more massive,
compact objects. However, while the CN over-abundances could be due to an
over-abundance of carbon, when the C-tracing G4300 index is plotted
against [MgFe]\arcmin in Figure~\ref{fig:g4300mgfe}, the NGC 5128 objects
fall closer to the model predictions, suggesting that the offset may in
fact be due to a difference in N abundance.

\subsection{Formation of Massive Compact Star Clusters}

Since UCDs were first discovered in the Virgo and Fornax clusters in the
past decade, studies of these compact stellar systems, in particular those
undertaken by \cite{mie06} and \cite{has05}, have come a long way in
classifying them consistently in terms of structural parameters. Recent
studies have garnered empirical evidence for a `break-point' between UCDs
and more classical GCs at ${\cal M}_{\rm dyn}\simeq2-3\times10^{6}\,
M_{\sun}$ and absolute magnitude $M_{V}\approx -11$ mag in that, above
this mass limit, $\Upsilon_{V}^{\rm dyn}$ increases and there are changes
in both the metallicity distributions and the size-luminosity relations
\citep{has05,kis06,mie06,mie08}. Various formation scenarios have been
proposed for these objects, the most popular being that they are either
the remnant cores of tidally stripped dE,Ns or the end results of merged
massive young clusters. Both scenarios have been supported by studies of
Virgo and Fornax cluster UCDs respectively \citep[e.g.][]{fel02, has05,
jon06, mie06}, although the idea that UCDs are simply the brightest, most
massive GCs remains the most straight-forward explanation
\citep[e.g.][]{mie04,mho04}.

\cite{hil09} discusses the implications of the various formation
scenarios. He suggests that a general formation scheme can be considered
in light of a mass break-point of a few $10^{6}\, M_{\sun}$; below this
mass limit clusters can be considered to be ``single-collapse'' globular
clusters (SCGCs), while above it the formation physics predict
``multiple-collapse'' globular clusters (MCGCs). In the case of SCGCs, the
present day objects would be expected to host a SSP comprised of stars
sharing both the same age and metallicity, i.e.~`classical' GCs, while for
MCGCs, the timescales of formation can vary from Myr to Gyr, thus,
allowing for chemically distinct multiple stellar generations with a range
of ages.

If the most massive GCs form within the most massive giant molecular
clouds (GMCs), and the most massive GMCs form in/around the most massive
galaxies (from purely statistical arguments), then it naturally follows
that the most massive clusters would be found preferentially near massive
galaxies such as NGC~5128, or near the central galaxies in clusters like
Fornax and Virgo.~If multiple SCGCs are formed through the fragmentation
of a massive GMC, then a MCGC would form through the merging of the SCGCs
on a timescale of $10\!-\!100$ Myr.~A MCGC formed through this channel
would result in a single-aged stellar population, probably with similar
metallicities \citep{hil09}.~On the other hand, if a timescale of Gyr is
invoked, the idea of stripped dE,Ns could be explained if the nuclear
clusters are formed through gradual merging of `classical' GCs, or if
there are periods of star formation triggered by gas being funneled to the
centre of a gas-rich galaxy. Regardless of what mechanism allowed the
massive clusters to grow, the evolution over a Hubble time in the tidal
field of a $10^{13} - 10^{15} M_\odot$ potential well can result in
objects with structural properties similar to UCDs.

The NGC 5128 clusters of this study show similarities to both Local Group
GCs and to the structural definition of UCDs. However, since no single
cluster shares all of the parameters that describe UCDs \citep[according
to][]{hil09} we hesitate to re-define any of them as such. Instead, these
objects serve as a sample that falls in the `grayest area' between the two
populations and as such are perfect tools to probe any chemodynamical
evolutionary connections between them.

The ages of the objects are indicated by Figure~\ref{fig:hbmgfe} to all be
old ($\gtrsim8\!-\!10$ Gyr, Fig.~\ref{fig:hbmgfe}), while
Figure~\ref{fig:mgbfemgfe} shows a range in [$\alpha$/Fe] ratios,
consistent with super-solar to solar values. The more metal-rich NGC 5128
clusters have metallicities similar to those of the nuclear clusters of
dE,Ns and for the most part share similar [$\alpha$/Fe] ratios. In
contrast, the [$\alpha$/Fe] ratios of most UCDs are significantly higher
and, therefore, exclude the possibility of dE,N cores as their
progenitors. These chemical features make NGC~5128 GCs consistent with the
idea that part of the sample may be stripped dE,Ns if the nuclear clusters
have formed through one of the mechanisms described above. The generally
older ages of the NGC 5128 objects compared to the dE,N cores also lend
weight to this interpretation. Interestingly, the three most metal-rich
clusters have super-solar [$\alpha$/Fe] and slightly younger ages than the
others, suggestive of shorter SFHs. These features are consistent with a
monolithic, fractional collapse and subsequent merging of massive GMCs. On
all counts, the least massive clusters of our GC sample show similarities
to the `classical' GCs in the Local Group and should probably be
considered as such.

\section{Conclusions}

In the past decade, massive compact objects with structural properties
similar to both Local Group globular clusters and nuclear star clusters of
dwarf galaxies have been found and studied extensively near the central
regions of massive galaxy clusters.~These objects show indications of
evolutionary pasts similar to the other compact systems and much
work has been done to investigate any connection between the various
populations. The nearby massive elliptical galaxy NGC 5128 provides an interesting laboratory to look for and study objects like these, primarily
because of its proximity \citep[$3.8\pm0.1$ Mpc, ][]{har09}, but also
because it is the central massive galaxy in a group and has undergone
(relatively) recent mergers. Consequently, it provides an environment that
could form UCDs with both of the current popular formation mechanisms:
stripping of galaxy cores and merging of star clusters.

In this work we study line of sight velocity dispersions for 23 massive
stellar systems in NGC 5128, derived from high-resolution spectra from two
observing programs on the 8.2 meter ESO/VLT telescope UT2 (Kueyen)
with UVES using the penalized pixel-fitting code of \cite{cap04}.~The
aperture corrected central velocity dispersions for a subsample of 21
clusters are used to derive dynamical mass estimates and are combined with
other structural parameters in order to compare them to Local Group GCs,
UCDs, and early-type galaxies. The star formation histories of the objects
are also investigated using chemical abundances derived from Lick/IDS indices.

The absence of dark matter in Local Group GCs is generally accepted
\citep[e.g.][]{mor96}. This paradigm is being challenged at the higher end
of the mass distribution in that clusters of mass
$\gtrsim2\!-\!3\times10^{6}\, M_{\sun}$ require either a top/bottom heavy
IMF or dark matter components to explain their higher mass-to-light
ratios. Our results support this finding in that, above this mass break,
the $\Upsilon_{V}^{\rm dyn}$ begins to increase significantly
$\Upsilon_{V}^{\rm dyn}\propto {\cal M}_{\rm dyn}^{0.24\pm0.02}$, see
Sect.~\ref{ln:fp}) and approaches those derived for UCDs found in the
Fornax and Virgo galaxy clusters. If one requires that UCDs share all of
the structural parameters described by \cite{hil09}, then no GCs of our
sample can be defined as such. However all but the lowest masses in the
sample are clearly transitional objects between the GC and UCD
populations, while UCDs themselves have been described as transitional
objects between GCs and early-type galaxies if an evolutionary connection
truly exists.

\cite{hil09} refers to UCDs as a ``mixed bag of objects'', but in light
of this study the expression may be even better suited to our sample of
objects. The lowest-mass clusters in our sample represent the
`classical' GCs of the Local Group in every aspect, without requiring
dark matter or any combination of strange IMF, composite stellar
populations or galactic interaction to explain their observed
parameters. Thus, we consider them to be typical single-collapse GCs.
The objects falling above $2\!-\!3\times10^{6}\, M_{\sun}$ confirm the
mass break reported by \cite{has05} in that their high mass-to-light
ratios begin to require non-baryonic components. Most of these objects
share metallicities similar to those of the dE,N cores; combined with
the ages and [$\alpha$/Fe] ratios, these objects support the notion that
they are the present-day remnants of dE,Ns that have been stripped of
their envelopes. Three of our NGC 5128 objects indicate short SFHs by
having super-solar [$\alpha$/Fe] ratios; combined with slightly lower
ages compared to the other objects, thus is suggestive of the merging of
single-collapse GCs, that possibly formed from a fragmented massive GMC.
Altogether, our results are consistent with the interpretation of
\cite{hil09} -- single vs.~multiple-collapse formation -- as well as
formation by the tidal stripping of dE,Ns in massive potential wells
\citep{zinn88, hil00, mey01, gne02, bek03, dri03, has07}. More
observations of transitional objects such as these in galaxy clusters
and massive galaxies like NGC 5128 will hopefully serve to further
constrain the formation mechanisms, and advance our understanding of the
evolutionary processes of massive, compact spheroidal systems.

\section{Acknowledgments}
We thank Patrick C\^{o}t\'{e} for providing measurements for some of the
UCD data prior to publication and for helpful discussions.~Many thanks go
out to Michele Cappellari for providing his pPXF code and valuable advice
on its usage.~We thank the anonymous referee for a constructive
report that helped to improve the paper. MAT acknowledges support through
the Co-Op program at the Herzberg Institute of Astrophysics.~THP
acknowledges support in form of a Plaskett Research Fellowship at the
Herzberg Institute of Astrophysics.~WEH and GLHH acknowledge support from
the Natural Sciences and Engineering Research Council of Canada.

\clearpage

\clearpage

\clearpage

\clearpage

\clearpage

\clearpage

\clearpage

\clearpage

\clearpage

\clearpage

\clearpage

\clearpage
\begin{deluxetable}{llrccccrr}
\tabletypesize{\scriptsize}
\tablewidth{0pt}
\tablecaption{Journal of Globular Cluster Observations \label{tab:gcobs}}
\tablehead{
\colhead{Cluster ID} & \colhead{Cluster ID} & \colhead{Exp. Time} & \colhead{Prog. ID} & \colhead{RA} & \colhead{DEC} & \colhead{V} & \colhead{S/N} & \colhead{S/N}\\
\colhead{[old]} & \colhead{[new]} & \colhead{[sec]} & \colhead{} & \colhead{[J2000]} & \colhead{[J2000]} & \colhead{[mag]}  & \colhead{[REDL]} & \colhead{[REDU]}
}
\startdata
f1.GC13    & GC0372  &	1800    & 69.D-0196 & 13 26 07.8 & $-42$ 51 59.7 & 18.35 & 4  & 4  \\
f1.GC16    & GC0378  &	1800    & 69.D-0196 & 13 26 10.6 & $-42$ 53 43.0 & 18.43 & 4  & 4  \\
f1.GC22    & GC0382  &	1800    & 69.D-0196 & 13 26 15.8 & $-42$ 55 01.0 & 18.09 & 5  & 5  \\
f1.GC23    & GC0397  &	1800    & 69.D-0196 & 13 26 23.8 & $-42$ 54 01.6 & 18.41 & 4  & 4  \\
f2.GC23    & GC0232  &	1800    & 69.D-0196 & 13 25 32.7 & $-43$ 07 02.1 & 18.77 & 3  & 4  \\
f2.GC26    & GC0218  &	1800    & 69.D-0196 & 13 25 30.5 & $-43$ 11 49.8 & 18.56 & 3  & 4  \\
f2.GC61    & GC0150  &	1800    & 69.D-0196 & 13 25 13.0 & $-43$ 07 59.9 & 18.20 & 3  & 4  \\
f2.GC81    & GC0123  & 	1800    & 69.D-0196 & 13 25 05.8 & $-43$ 10 30.4 & 17.74 & 3  & 4  \\
HCH99-C2   & GC0171  &	1200    & 69.D-0196 & 13 25 16.4 & $-43$ 02 06.7 & 18.21 & 3  & 3  \\
HCH99-C15  & GC0213  &	1200    & 69.D-0196 & 13 25 29.9 & $-43$ 00 03.9 & 17.56 & 3  & 4  \\
HCH99-C16  & GC0217  &	1800    & 69.D-0196 & 13 25 30.1 & $-42$ 59 35.4 & 18.45 & 2  & 2  \\
HCH99-C18  & GC0225  &	1200    & 69.D-0196 & 13 25 31.6 & $-43$ 00 01.8 & 17.07 & 5  & 6  \\
HCH99-C21  & GC0242  &	1200    & 69.D-0196 & 13 25 34.4 & $-43$ 03 26.5 & 18.41 & 2  & 2  \\
HGHH92-C1  & GC0005  &	2400    & 69.D-0196 & 13 23 44.0 & $-43$ 11 13.5 & 17.42 & 3  & 3  \\
HGHH92-C7  & GC0365  &	18000   & 69.D-0094 & 13 26 05.2 & $-42$ 56 32.9 & 17.17 & 8  & 12 \\
HGHH92-C11 & GC0077  &	2400    & 69.D-0196 & 13 24 54.6 & $-43$ 01 22.8 & 17.91 & 2  & 3  \\
HGHH92-C17 & GC0265  &	23400   & 69.D-0094 & 13 25 39.6 & $-42$ 55 58.2 & 17.63 & 11 & 11 \\
HGHH92-C21 & GC0320  &	1200    & 69.D-0196 & 13 25 52.7 & $-43$ 05 46.6 & 17.87 & 3  & 3  \\
HGHH92-C22 & GC0326  &	1800    & 69.D-0196 & 13 25 53.3 & $-42$ 59 07.6 & 18.15 & 4  & 5  \\
HGHH92-C23 & GC0330  &	23400   & 69.D-0094 & 13 25 54.3 & $-42$ 59 25.4 & 17.22 & 10 & 11 \\
HGHH92-C29 & GC0041  &	1200    & 69.D-0196 & 13 24 40.4 & $-43$ 18 03.9 & 18.15 & 3  & 4  \\
HGHH92-C41 & GC0040  &	1800    & 69.D-0196 & 13 24 38.9 & $-43$ 20 05.6 & 18.59 & 3  & 5  \\
HGHH92-C44 & GC0227  &	3600    & 69.D-0196 & 13 25 31.8 & $-43$ 19 23.6 & 18.69 & 3  & 4  \\
\enddata
\tablecomments{The listed targets are globular clusters in NGC~5128 that are part of our sample. Col. (1) lists the cluster ID according to the literature sources, col. (2) lists the new, homogeneous IDs of \cite{woo07}, col. (3) lists the total integration time in seconds, col. (4) lists the program IDs for the original spectra and cols. (5) and (6) list the right ascension and declination in J2000 coordinates. Cols. (7) shows the apparent cluster V magnitudes not corrected for foreground reddening, while cols. (8) and (9) show the average S/N per pixel over the spectral regions used.}
\end{deluxetable}

\clearpage
\begin{deluxetable}{lcccccrr}
\tabletypesize{\footnotesize}
\tablewidth{0pt}
\tablecaption{Journal of Template Star Observations \label{tab:stars}}
\tablehead{
\colhead{Star} & \colhead{Spec. Type} & \colhead{Exp. Time} & \colhead{Prog. ID} & \colhead{RA} & \colhead{DEC} & \colhead{S/N} & \colhead{S/N} \\ 
\colhead{} & \colhead{} & \colhead{[sec]} & \colhead{} & \colhead{[J2000]} & \colhead{[J2000]} & \colhead{[REDL]} & \colhead{[REDU]}
}
\startdata
HD33771  & G0       &  165 & 69.D-0196 & 05 10 49.7 & $-$37 49 04.1 & 10 & 19 \\
HD66141  & K2III    &  10  & 69.D-0196 & 08 02 16.0 & +02 20 07.1   & 4  & 7  \\
HD81797  & K3II-III &  5   & 69.D-0196 & 09 27 35.2 & $-$08 39 29.2 & 3  & 5  \\
HD93529  & G7w      &  120 & 69.D-0196 & 10 47 29.1 & $-$25 26 13.4 & 8  & 15 \\
HD103295 & G5/6III  &  75  & 69.D-0196 & 11 53 26.9 & $-$28 38 12.0 & 8  & 15 \\
HD146051 & M0.5III  &  2   & 69.D-0196 & 16 14 20.6 & +03 54 32.0   & 3  & 5  \\
HD150798 & K2II-III &  1   & 69.D-0196 & 16 48 39.6 & $-$69 01 40.8 & 3  & 5  \\
HD161096 & K2III    &  1   & 69.D-0196 & 17 43 28.4 & +04 34 03.1   & 3  & 6  \\
HD165195 & K3p	    &  25  & 69.D-0196 & 18 04 39.9 & +03 46 47.2   & 7  & 14 \\
HD168454 & K3IIIa   &  1   & 69.D-0196 & 18 20 59.7 & $-$29 49 40.2 & 3  & 6  \\
HD171391 & G8III    &  10  & 69.D-0196 & 18 35 02.2 & $-$10 58 39.3 & 5  & 9  \\
HD196983 & K2III    &  100 & 69.D-0196 & 20 41 50.6 & $-$33 53 18.4 & 4  & 6  \\
HD203638 & K0III    &  30  & 69.D-0196 & 21 24 09.4 & $-$20 51 07.8 & 4  & 6  \\
HR4614   & G3Ib     &  2   & 69.D-0094 & 12 06 20.5 & $-$68 39 07.8 & 4  & 7  \\
HR5293   & K4II     &  4   & 69.D-0094 & 14 10 31.1 & $-$70 18 18.5 & 3  & 5  \\
HR5461   & K0.5II   &  2   & 69.D-0094 & 14 40 33.1 & $-$56 26 28.5 & 4  & 8  \\
HR5547   & G8II     &  2   & 69.D-0094 & 15 00 11.0 & $-$77 09 40.7 & 5  & 9  \\
HR5891   & G5IIa    &  1   & 69.D-0094 & 15 55 29.9 & $-$68 36 09.9 & 4  & 7  \\
HR6856   & K2II     &  2   & 69.D-0094 & 18 20 55.2 & $-$37 29 17.4 & 3  & 6  \\
HR7277   & K0Iab    &  2   & 69.D-0094 & 19 13 13.6 & $-$25 54 23.0 & 3  & 6  \\
HR7586   & K4II     &  3   & 69.D-0094 & 19 56 57.6 & $-$57 55 33.3 & 3  & 5  \\
\enddata
\tablecomments{Template stars used in the velocity dispersion measurements. Col. (1) lists the star IDs and col. (2) lists the spectral type. Cols. (3-6) lists the exposure times, program ID, right ascensions and declinations while cols. (7) and (8) give the average S/N per pixel over the spectral regions used.}
\end{deluxetable}

\clearpage
\begin{deluxetable}{lccccc}
\tabletypesize{\small}
\tablewidth{0pt}
\tablecaption{Comparison of Measured and Modeled Velocity Dispersion \label{tbl:sigcomp}}
\tablehead{
\colhead{Object} & \colhead{$v_{r}$} &\colhead{$\sigma_{\rm ppxf}$} & \colhead{$\sigma_{m,obs}$} & \colhead{$\sigma_{m,global}$} & \colhead{$\sigma_{o}$} \\ 
\colhead{} & \colhead{[km s$^{-1}$]} & \colhead{[kms$^{-1}$]} & \colhead{[kms$^{-1}$]} & \colhead{[kms$^{-1}$]} & \colhead{[kms$^{-1}$]}
}
\startdata
GC0372  & 699$\pm$10      &  16.5$\pm$1.8  & 16.6$^{+2.0}_{-2.2}$ & 15.4$^{+1.9}_{-2.1}$ & 20.2$^{+2.5}_{-2.7}$ \\
GC0378  & 608$\pm$10      &  14.2$\pm$1.3  & 14.2$^{+2.2}_{-1.8}$ & 12.8$^{+2.0}_{-1.7}$ & 16.4$^{+2.6}_{-2.1}$ \\
GC0382  & 582$\pm$10      &  14.3$\pm$3.2  & 14.3$^{+3.2}_{-3.2}$ & 14.0$^{+3.2}_{-3.1}$ & 17.5$^{+4.5}_{-4.0}$ \\
GC0397  & 402$\pm$10      &  15.6$\pm$1.4  & 15.8$^{+1.8}_{-1.4}$ & 14.6$^{+1.7}_{-1.3}$ & 19.5$^{+2.2}_{-1.8}$ \\
GC0232  & 776$\pm$10      &  14.1$\pm$2.1  & 14.0$^{+2.7}_{-2.2}$ & 13.1$^{+2.6}_{-2.0}$ & 17.2$^{+3.4}_{-2.7}$ \\
GC0218  & 640$\pm$10      &  12.7$\pm$1.6  & 12.7$^{+1.8}_{-1.6}$ & 11.4$^{+1.6}_{-1.4}$ & 15.0$^{+2.1}_{-1.9}$ \\
GC0150  & 612$\pm$10      &  15.9$\pm$1.2  & 16.1$^{+1.7}_{-1.1}$ & 15.2$^{+1.7}_{-1.0}$ & 19.6$^{+2.1}_{-1.3}$ \\
GC0123  & 436$\pm$10      &  14.9$\pm$1.4  & 14.9$^{+1.9}_{-1.5}$ & 12.4$^{+1.6}_{-1.3}$ & 17.4$^{+2.2}_{-1.8}$ \\
GC0171  & 296$\pm$10      &  16.3$\pm$1.8  & 16.1$^{+1.2}_{-2.6}$ & 14.1$^{+1.0}_{-2.2}$ & 18.8$^{+1.4}_{-3.0}$ \\
GC0213  & 519$\pm$10      &  27.5$\pm$3.2  & 27.6$^{+3.4}_{-4.9}$ & 26.4$^{+3.3}_{-4.7}$ & 33.4$^{+4.1}_{-5.9}$ \\
GC0217  & 459$\pm$11      &  10.9$\pm$6.3  & 10.9$^{+7.5}_{-6.1}$ & 9.3$^{+6.3}_{-5.2}$  & 11.3$^{+7.7}_{-6.3}$ \\
GC0225  & 452$\pm$10      &  22.0$\pm$1.2  & 21.7$^{+1.1}_{-1.1}$ & 19.1$^{+1.0}_{-1.0}$ & 24.1$^{+1.2}_{-1.3}$ \\
GC0242  & 661$\pm$10      &  12.1$\pm$1.4  & 12.1$^{+2.4}_{-1.3}$ & 11.2$^{+2.2}_{-1.2}$ & 13.3$^{+2.6}_{-1.4}$ \\
GC0005  & 645$\pm$10      &  15.9$\pm$1.6  & \nodata         & \nodata        & \nodata       \\
GC0365  & 594$\pm$11      &  23.7$\pm$2.4  & 23.8$^{+2.6}_{-2.7}$ & 21.6$^{+2.4}_{-2.5}$ & 27.3$^{+3.0}_{-3.1}$ \\
GC0077  & 749$\pm$10      &  16.7$\pm$1.5  & 16.7$^{+1.7}_{-2.3}$ & 14.7$^{+1.5}_{-2.0}$ & 19.5$^{+2.0}_{-2.7}$ \\
GC0265  & 787$\pm$10      &  20.8$\pm$2.9  & 20.9$^{+3.1}_{-1.2}$ & 19.8$^{+2.9}_{-1.2}$ & 24.1$^{+3.5}_{-1.4}$ \\
GC0320  & 455$\pm$10      &  20.0$\pm$1.4  & 19.9$^{+1.3}_{-1.8}$ & 17.7$^{+1.1}_{-1.6}$ & 23.0$^{+1.5}_{-2.0}$ \\
GC0326  & 575$\pm$10      &  19.4$\pm$1.8  & 19.4$^{+1.8}_{-2.0}$ & 18.4$^{+1.7}_{-1.9}$ & 23.1$^{+2.1}_{-2.4}$ \\
GC0330  & 669$\pm$10      &  41.5$\pm$3.7  & 41.6$^{+4.3}_{-4.6}$ & 39.8$^{+4.1}_{-4.4}$ & 50.0$^{+5.2}_{-5.5}$ \\
GC0041  & 724$\pm$10      &  17.6$\pm$1.8  & 17.6$^{+2.7}_{-1.4}$ & 15.4$^{+2.3}_{-1.2}$ & 19.7$^{+3.0}_{-1.5}$ \\
GC0040  & 361$\pm$10      &  13.7$\pm$1.3  & 13.7$^{+2.1}_{-0.9}$ & 12.4$^{+1.9}_{-0.8}$ & 15.7$^{+2.4}_{-1.0}$ \\
GC0227  & 496$\pm$11      &  13.6$\pm$1.2  & 13.5$^{+1.4}_{-1.4}$ & 12.5$^{+1.3}_{-1.3}$ & 15.7$^{+1.6}_{-1.7}$ \\
\enddata
\tablecomments{Radial velocities measured with \emph{fxcor} with $\sigma$'s measured by pPXF and modeled by the aperture correction code of \cite{hil07}. Col. (2) lists $v_{r}$, while col. (3) shows the observed $\sigma$ measured with pPXF and col. (4) shows the observed $\sigma$ predicted by the code for comparison. The other two columns show the global and central $\sigma$'s predicted by the aperture correction code.}
\end{deluxetable}

\clearpage
\begin{deluxetable}{lccccrr}
\tabletypesize{\scriptsize}
\tablecaption{Photometric Star Cluster Parameters\label{tbl:phot}}
\tablehead{
\colhead{Cluster ID} &  \colhead{$V_{0}$} & \colhead{$(U-B)_{0}$} & \colhead{$(B-V)_{0}$} & \colhead{$(V-I)_{0}$} & \colhead{$M_{V}$} & \colhead{$Z$} \\
\colhead{[new]} & \colhead{[mag]} & \colhead{[mag]} & \colhead{[mag]} & \colhead{[mag]} & \colhead{[mag]} & \colhead{}
}
\startdata
GC0372 & 18.04 & 0.07$\pm$0.04 & 0.73$\pm$0.01 & 0.85$\pm$0.03 & $-$9.70  & 0.0012  \\
GC0378 & 18.09 & 0.30$\pm$0.04 & 0.84$\pm$0.01 & 0.99$\pm$0.03 & $-$9.62  & 0.0062   \\
GC0382 & \nodata & \nodata     & \nodata       & \nodata       & $-$9.96 & \nodata  \\
GC0397 & 18.10 & 0.21$\pm$0.04 & 0.78$\pm$0.01 & 0.91$\pm$0.03 & $-$9.64  & \nodata  \\
GC0232 & 18.43 & 0.23$\pm$0.07 & 0.80$\pm$0.02 & 0.95$\pm$0.03 & $-$9.28  & 0.0034  \\
GC0218 & 18.19 & 0.35$\pm$0.06 & 0.89$\pm$0.02 & 1.03$\pm$0.03 & $-$9.49  & 0.0065   \\
GC0150 & 17.83 & 0.31$\pm$0.05 & 0.83$\pm$0.01 & 0.99$\pm$0.03 & $-$9.85 & 0.0058  \\
GC0123 & 17.40 & \nodata       & \nodata       & \nodata       & $-$10.31 & \nodata  \\
GC0171 & 17.87 & 0.22$\pm$1.01 & 0.74$\pm$0.27 & 0.84$\pm$0.09 & $-$10.12 & \nodata   \\
GC0213 & 17.42 & \nodata       & \nodata       & 1.06$\pm$0.08 & $-$10.61 & \nodata  \\
GC0217 & 18.13 & \nodata       & \nodata       & 0.79$\pm$0.08 & $-$9.87 & 0.0006   \\
GC0225 & 16.83 & $-$0.11$\pm$2.54 & 0.89$\pm$0.78    & 0.89$\pm$0.22    & $-$11.17 & \nodata    \\
GC0242 & 17.91 & \nodata       & \nodata       & 0.78$\pm$0.08 & $-$10.07 & 0.0060   \\
GC0005 & \nodata & \nodata     & \nodata       & \nodata       & \nodata  & 0.0008  \\
GC0365 & 16.83 & 0.17$\pm$0.03 & 0.75$\pm$0.01 & 0.91$\pm$0.03 & $-$10.88 & 0.0025   \\
GC0077 & 17.54 & 0.39$\pm$0.05 & 0.94$\pm$0.01 & 1.12$\pm$0.03 & $-$10.14 & 0.0088  \\
GC0265 & 17.29 & 0.14$\pm$0.04 & 0.77$\pm$0.01 & 0.88$\pm$0.03 & $-$10.42 & 0.0030  \\
GC0320 & 17.53 & 0.22$\pm$0.04 & 0.78$\pm$0.01 & 0.92$\pm$0.03 & $-$10.18 & 0.0025  \\
GC0326 & 17.81 & 0.16$\pm$0.05 & 0.79$\pm$0.02 & 0.91$\pm$0.03 & $-$9.90 & 0.0020  \\
GC0330 & 16.88 & 0.45$\pm$0.04 & 0.96$\pm$0.01 & 1.10$\pm$0.03 & $-$11.45 & 0.0079  \\
GC0041 & 17.75 & 0.41$\pm$0.04 & 0.89$\pm$0.01 & 1.08$\pm$0.03 & $-$9.90 & 0.0092  \\
GC0040 & 18.19 & 0.40$\pm$0.06 & 0.89$\pm$0.02 & 1.09$\pm$0.03 & $-$9.46  & 0.0077   \\
GC0227 & 18.32 & 0.06$\pm$0.04 & 0.69$\pm$0.01 & 0.85$\pm$0.03 & $-$9.36  & 0.0009   \\
\enddata
\tablecomments{The stellar clusters observed. Col.~(1) lists the cluster ID, while the de-reddened apparent $V$ magnitudes are listed in col. (2) based on $R_{V}$=3.1, using $E_{B-V}$ quoted from \cite{pen04}. Values for GCs 0171, 0213, 0217, 0225 and 0242 were quoted from \cite{hol99}. Cols.~(3-5) show the de-reddened $(U\!-\!B)_0$, $(B\!-\!V)_0$, and $(V\!-\!I)_0$ color indices (where available) and col. (6) are the absolute $V$-band magnitudes, while col. (7) shows the $Z$ values from \cite{bea08}. The uncertainty of $M_{V}$ is $\pm$0.05 mag based on $(m\!-\!M)_{0}$=27.88$\pm$0.05 (see \S~\ref{ln:sp}).}
\end{deluxetable}

\clearpage
\begin{deluxetable}{lcccc}
\tabletypesize{\small}
\tablewidth{0pt}
\tablecaption{Photometric Cluster Properties \label{tbl:photML}}
\tablehead{
\colhead{ID} & \colhead{$\log t_{\rm Salp}$} & \colhead{$\Upsilon^{\rm phot}_{\rm Salp}$} & \colhead{$\log t_{\rm Chab}$} & \colhead{$\Upsilon^{\rm phot}_{\rm Chab}$} \\ \colhead{} & \colhead{[Gyr]} & \colhead{$[M_{\sun}/L_{\sun}]$} & \colhead{[Gyr]} & \colhead{$[M_{\sun}/L_{\sun}]$}
}
\startdata
GC0372	& 9.9$_{-0.4}^{+0.4}$	& 2.7$_{-1.5}^{+2.0}$	& 9.9$_{-0.4}^{+0.4}$	& 1.4$_{-0.6}^{+1.6}$\\
GC0378	& 9.7$_{-0.3}^{+0.2}$	& 2.5$_{-1.1}^{+1.2}$	& 9.7$_{-0.3}^{+0.3}$	& 2.2$_{-1.3}^{+0.1}$\\
GC0232	& 9.9$_{-0.4}^{+0.4}$	& 2.9$_{-1.3}^{+2.7}$	& 9.9$_{-0.3}^{+0.4}$	& 1.9$_{-0.9}^{+1.5}$\\
GC0218	& 9.9$_{-0.3}^{+0.4}$	& 3.7$_{-1.3}^{+2.9}$	& 10.0$_{-0.3}^{+0.3}$	& 2.3$_{-0.9}^{+1.7}$\\
GC0150	& 9.7$_{-0.3}^{+0.2}$	& 2.5$_{-1.1}^{+1.1}$	& 9.8$_{-0.3}^{+0.2}$	& 1.5$_{-0.6}^{+0.7}$\\
GC0123	& \nodata		& \nodata		& \nodata		& 3.0$_{-0.0}^{+0.0}$\\
GC0217	& 9.3$^{*}$		& 0.8$^{*}$		& 9.4$^{*}$		& 0.6$^{*}$            \\
GC0242	& 9.1$^{*}$		& 0.6$^{*}$		& 9.1$^{*}$		& 0.4$^{*}$		\\
GC0365	& 9.8$_{-0.3}^{+0.2}$	& 2.2$_{-0.8}^{+0.9}$	& 9.9$_{-0.3}^{+0.4}$	& 1.6$_{-0.7}^{+1.7}$\\
GC0077	& 10.1$_{-0.2}^{+0.2}$	& 4.7$_{-1.1}^{+2.3}$	& 10.0$_{-0.1}^{+0.3}$	& 2.7$_{-0.5}^{+1.7}$\\
GC0265	& 9.5$_{-0.2}^{+0.5}$	& 1.5$_{-0.4}^{+1.8}$	& 9.6$_{-0.4}^{+0.7}$	& 1.0$_{-0.5}^{+2.3}$\\
GC0320	& 10.0$_{-0.4}^{+0.3}$	& 3.1$_{-1.6}^{+2.1}$	& 10.0$_{-0.4}^{+0.2}$	& 2.1$_{-1.1}^{+1.2}$\\
GC0326	& 9.9$_{-0.4}^{+0.4}$	& 2.8$_{-1.3}^{+2.2}$	& 10.0$_{-0.3}^{+0.3}$	& 1.7$_{-0.8}^{+1.4}$\\
GC0330	& 10.0$_{-0.3}^{+0.3}$	& 4.7$_{-1.9}^{+2.3}$	& 10.1$_{-0.2}^{+0.2}$	& 2.9$_{-1.0}^{+1.3}$\\
GC0041	& 10.0$_{-0.3}^{+0.2}$	& 4.3$_{-1.6}^{+1.7}$	& 9.9$_{-0.2}^{+0.2}$	& 2.2$_{-0.6}^{+0.7}$\\
GC0040	& 9.9$_{-0.2}^{+0.2}$	& 3.8$_{-1.1}^{+1.2}$	& 10.0$_{-0.2}^{+0.2}$	& 2.4$_{-0.7}^{+1.2}$\\
GC0227	& 9.9$_{-0.3}^{+0.4}$	& 2.3$_{-1.0}^{+1.8}$	& 10.0$_{-0.3}^{+0.4}$	& 1.5$_{-0.7}^{+1.4}$\\
\enddata
\tablecomments{Clusters marked with $^{*}$ only had $(V-I)_{0}$ data available, and so the age and $\Upsilon$ are expected to be uncertain. Cluster GC0123 had a photometric $\Upsilon$ listed in \cite{mcl08} who used a Wilson model fit with a Chabrier (2003) disk-IMF.}
\end{deluxetable}

\clearpage
\begin{deluxetable}{lccccccccr}
\tabletypesize{\scriptsize}
\tablewidth{0pt}
\tablecaption{Dynamical and Structural Cluster Parameters \label{tab:ds}}
\tablehead{
\colhead{ID} & \colhead{$\sigma_{\rm ppxf}$} & \colhead{$\sigma_{\rm Rej07}$} & \colhead{$\sigma_{o}$} & \colhead{$M_{dyn}$} & \colhead{$\Upsilon_{V}^{dyn}$} & \colhead{$\Upsilon_{V}^{phot}$} & \colhead{$R_{h}$} & \colhead{r$_{c}$} & \colhead{c} \\ \colhead{} & \colhead{[km s$^{-1}$]}  & \colhead{[km s$^{-1}$]} & \colhead{[km s$^{-1}$]} & \colhead{[$\times$$10^{6}$ $M_{\sun}$]} & \colhead{[$M_{\sun}$/$L_{\sun}$]} & \colhead{[$M_{\sun}$/$L_{\sun}$]} & \colhead{[pc]} & \colhead{[pc]} & \colhead{} 
}
\startdata
GC0372  & 16.5$\pm$1.8 & 15.7$\pm$1.9 & 20.2$^{+2.5}_{-2.7}$  & 2.5$^{+0.9}_{-0.9}$ & 3.3$^{+1.2}_{-1.2}$ & 1.4$^{+1.6}_{-0.6}$    & 3.54$\pm$0.348  & 0.68$\pm$0.297 & 2.00 \\
GC0378  & 14.2$\pm$1.3 & 12.6$\pm$0.8 & 16.4$^{+2.6}_{-2.1}$  & 1.5$^{+0.5}_{-0.5}$ & 2.2$^{+0.8}_{-0.6}$ & 2.2$^{+0.1}_{-1.3}$    & 3.26$\pm$0.131  & 0.58$\pm$0.109 & 1.87 \\
GC0382  & 14.3$\pm$3.2 & 4.9$\pm$1.1  & 17.5$^{+4.5}_{-4.0}$  & 1.1$^{+0.8}_{-0.7}$ & \nodata             & \nodata                & 2.14$\pm$0.315  & 0.58$\pm$0.297 & 1.70 \\
GC0397  & 15.6$\pm$1.4 & 14.2$\pm$1.1 & 19.5$^{+2.2}_{-1.8}$  & 1.8$^{+0.6}_{-0.5}$ & 2.6$^{+0.9}_{-0.7}$ & \nodata                & 2.73$\pm$0.263  & 0.50$\pm$0.198 & 2.20 \\
GC0232  & 14.1$\pm$2.1 & 14.4$\pm$1.5 & 17.2$^{+3.4}_{-2.7}$  & 1.3$^{+0.7}_{-0.6}$ & 2.6$^{+1.3}_{-1.1}$ & 1.9$^{+1.5}_{-0.9}$    & 2.58$\pm$0.324  & 0.50$\pm$0.394 & 2.00 \\
GC0218  & 12.7$\pm$1.6 & 11.1$\pm$0.7 & 15.0$^{+2.1}_{-1.9}$  & 2.1$^{+0.9}_{-0.8}$ & 3.2$^{+1.3}_{-1.2}$ & 2.3$^{+1.7}_{-0.9}$    & 5.23$\pm$0.718  & 1.18$\pm$0.206 & 1.90 \\
GC0150  & 15.9$\pm$1.2 & 14.6$\pm$0.7 & 19.6$^{+2.1}_{-1.3}$  & 1.3$^{+0.5}_{-0.4}$ & 3.2$^{+0.6}_{-0.5}$ & 1.5$^{+0.7}_{-0.6}$    & 1.92$\pm$0.383  & 0.40$\pm$0.197 & 2.00 \\
GC0123  & 14.9$\pm$1.4 & 13.1$\pm$0.5 & 17.4$^{+2.2}_{-1.8}$  & 5.6$^{+2.2}_{-2.0}$ & 3.0$^{+0.6}_{-0.5}$ & 3.0$^{+0.0}_{-0.0}$    & 10.60$\pm$1.51  & 1.18$\pm$0.206 & 2.40 \\
GC0171  & 16.3$\pm$1.8 & 14.1$\pm$0.5 & 18.8$^{+1.4}_{-3.0}$  & 6.9$^{+1.7}_{-3.0}$ & 8.0$^{+2.0}_{-3.5}$ & \nodata                & 11.19$\pm$1.05  & 0.97$\pm$0.105 & 1.50 \\
GC0213  & 27.5$\pm$3.2 & 21.3$\pm$1.7 & 33.4$^{+4.1}_{-5.9}$  & 11.2$^{+3.1}_{-4.3}$& 7.1$^{+2.0}_{-3.5}$ & \nodata                & 5.74$\pm$0.170  & 1.53$\pm$0.104 & 1.00 \\
GC0217  & 10.9$\pm$6.3 & 9.5$\pm$1.4  & 11.3$^{+7.7}_{-6.3}$  & 2.6$^{+3.7}_{-3.1}$ & 3.8$^{+5.4}_{-4.4}$ & 0.6                    & 11.89$\pm$0.520 & 0.80$\pm$0.159 & 1.60 \\
GC0225  & 22.0$\pm$1.2 & 21.2$\pm$1.1 & 24.1$^{+1.2}_{-1.3}$  & 13.6$^{+1.5}_{-1.6}$& 5.5$^{+0.6}_{-0.7}$ & \nodata                & 13.46$\pm$0.162 & 1.14$\pm$0.014 & 1.50 \\
GC0242  & 12.1$\pm$1.4 & 10.6$\pm$2.3 & 13.3$^{+2.6}_{-1.4}$  & 2.2$^{+1.6}_{-1.2}$ & 3.0$^{+2.3}_{-1.7}$ & 0.4                    & 7.00$\pm$2.52   & 2.52$\pm$1.02  & 0.80 \\
GC0005  & 15.9$\pm$1.6 & 12.9$\pm$0.8 & \nodata    	      & \nodata    	    & \nodata      	  & \nodata                & \nodata         & \nodata      & \nodata \\
GC0365  & 23.7$\pm$2.4 & 21.1$\pm$0.1 & 27.3$^{+3.0}_{-3.1}$  & 9.6$^{+2.2}_{-2.3}$ & 4.3$^{+1.0}_{-1.0}$ & 1.6$^{+1.7}_{-0.7}$    & 7.36$\pm$0.098  & 1.39$\pm$0.108 & 1.83 \\
GC0077  & 16.7$\pm$1.5 & 19.2$\pm$0.4 & 19.5$^{+2.0}_{-2.7}$  & 5.1$^{+1.1}_{-1.5}$ & 4.3$^{+0.9}_{-1.3}$ & 2.7$^{+1.7}_{-0.5}$    & 7.66$\pm$0.095  & 1.28$\pm$0.108 & 1.88 \\
GC0265  & 20.8$\pm$2.9 & 20.9$\pm$1.6 & 24.1$^{+3.5}_{-1.4}$  & 5.7$^{+1.8}_{-0.8}$ & 3.9$^{+1.2}_{-0.5}$ & 1.0$^{+2.3}_{-0.5}$    & 5.59$\pm$0.117  & 2.23$\pm$0.106 & 1.43 \\
GC0320  & 20.0$\pm$1.4 & 19.0$\pm$0.1 & 23.0$^{+1.5}_{-2.0}$  & 6.3$^{+0.9}_{-1.2}$ & 5.3$^{+0.8}_{-1.0}$ & 2.1$^{+1.2}_{-1.1}$    & 6.83$\pm$0.105  & 1.19$\pm$0.109 & 1.86 \\
GC0326  & 19.4$\pm$1.8 & 17.9$\pm$0.1 & 23.1$^{+2.1}_{-2.4}$  & 3.5$^{+0.8}_{-0.8}$ & 3.8$^{+0.8}_{-0.9}$ & 1.7$^{+1.4}_{-0.8}$    & 3.73$\pm$0.128  & 1.08$\pm$0.109 & 1.62 \\
GC0330  & 41.5$\pm$3.7 & 30.5$\pm$0.2 & 50.0$^{+5.2}_{-5.5}$  & 14.2$^{+3.5}_{-3.7}$& 6.6$^{+1.6}_{-1.7}$ & 2.9$^{+1.3}_{-1.0}$    & 3.25$\pm$0.131  & 0.86$\pm$0.109 & 1.67 \\
GC0041  & 17.6$\pm$1.8 & 16.1$\pm$0.8 & 19.7$^{+3.0}_{-1.5}$  & 4.6$^{+1.5}_{-0.8}$ & 4.7$^{+1.5}_{-0.8}$ & 2.2$^{+0.7}_{-0.6}$    & 6.75$\pm$0.106  & 1.17$\pm$0.109 & 1.87 \\
GC0040  & 13.7$\pm$1.3 & 11.5$\pm$1.3 & 15.7$^{+2.4}_{-1.0}$  & 1.9$^{+0.6}_{-0.3}$ & 3.0$^{+1.0}_{-0.5}$ & 2.4$^{+1.2}_{-0.7}$    & 4.42$\pm$0.131  & 0.77$\pm$0.109 & 1.87 \\
GC0227  & 13.6$\pm$1.2 & 13.1$\pm$1.0 & 15.7$^{+1.6}_{-1.7}$  & 2.4$^{+0.5}_{-0.6}$ & 4.2$^{+0.9}_{-1.0}$ & 1.5$^{+1.4}_{-0.7}$    & 5.59$\pm$0.117  & 1.22$\pm$0.108 & 1.70 \\
\enddata
\tablecomments{Structural Parameters for the observed stellar clusters. Cluster IDs are shown in col. (1), col. (2) lists $\sigma$ measured by pPXF, col. (3) lists the $\sigma$ values recorded by \cite{rej07} for comparison, while col. (4) shows $\sigma_{o}$ which was used for all relevant calculations. The virial masses and dynamical/photometric $\Upsilon_{V}$ are listed in cols. (5-7). Cols. (8), (9) and (10) list the projected half-light radii, ${\rm R}_{h}$, core radii, $r_{c}$, and concentration parameters, $c$, respectively and were taken from \cite{har02} with $c$ having an uncertainty of $\pm0.15$. }
\end{deluxetable}

\clearpage
\begin{deluxetable}{lccc}
\tabletypesize{\small}
\tablewidth{0pt}
\tablecaption{$\hat{\kappa}$-Space Parameters \label{tbl:kap}}
\tablehead{
\colhead{Object} & \colhead{$\hat{\kappa}_{1}$} & \colhead{$\hat{\kappa}_{2}$} & \colhead{$\hat{\kappa}_{3}$}
}
\startdata
GC0372	&	0.113$^{+0.082}_{-0.087}$	&	5.32$^{+0.063}_{-0.066}$	&	0.624$^{+0.138}_{-0.140}$ \\
GC0378	&	-0.040$^{+0.098}_{-0.080}$	&	5.29$^{+0.059}_{-0.049}$	&	0.515$^{+0.094}_{-0.082}$ \\	
GC0397	&	0.011$^{+0.075}_{-0.064}$	&	5.50$^{+0.060}_{-0.056}$	&	0.565$^{+0.133}_{-0.129}$ \\
GC0232	&	-0.083$^{+0.127}_{-0.104}$	&	5.41$^{+0.091}_{-0.081}$	&	0.559$^{+0.186}_{-0.176}$ \\
GC0218	&	0.050$^{+0.096}_{-0.088}$	&	4.81$^{+0.081}_{-0.078}$	&	0.612$^{+0.186}_{-0.183}$ \\
GC0150	&	-0.093$^{+0.089}_{-0.074}$	&	5.91$^{+0.101}_{-0.096}$	&	0.412$^{+0.256}_{-0.252}$ \\
GC0123	&	0.358$^{+0.089}_{-0.077}$	&	4.61$^{+0.080}_{-0.076}$	&	0.600$^{+0.189}_{-0.186}$ \\
GC0171	&	0.422$^{+0.054}_{-0.102}$	&	4.32$^{+0.051}_{-0.071}$	&	0.841$^{+0.132}_{-0.142}$ \\
GC0213	&	0.570$^{+0.076}_{-0.109}$	&	5.33$^{+0.046}_{-0.064}$	&	0.813$^{+0.072}_{-0.096}$ \\
GC0217	&	0.128$^{+0.419}_{-0.343}$	&	4.01$^{+0.242}_{-0.199}$	&	0.657$^{+0.346}_{-0.285}$ \\
GC0225	&	0.632$^{+0.031}_{-0.033}$	&	4.62$^{+0.019}_{-0.020}$	&	0.749$^{+0.029}_{-0.031}$ \\ 
GC0242	&	0.066$^{+0.163}_{-0.128}$	&	4.55$^{+0.182}_{-0.172}$	&	0.597$^{+0.461}_{-0.454}$ \\
GC0365	&	0.523$^{+0.068}_{-0.070}$	&	5.16$^{+0.039}_{-0.041}$	&	0.684$^{+0.058}_{-0.059}$ \\
GC0077	&	0.328$^{+0.063}_{-0.085}$	&	4.78$^{+0.037}_{-0.049}$	&	0.689$^{+0.054}_{-0.071}$ \\
GC0265	&	0.362$^{+0.089}_{-0.036}$	&	5.21$^{+0.052}_{-0.023}$	&	0.659$^{+0.077}_{-0.039}$ \\
GC0320	&	0.394$^{+0.040}_{-0.054}$	&	4.94$^{+0.024}_{-0.032}$	&	0.741$^{+0.038}_{-0.048}$ \\
GC0326	&	0.211$^{+0.057}_{-0.065}$	&	5.39$^{+0.036}_{-0.040}$	&	0.656$^{+0.063}_{-0.068}$ \\
GC0330	&	0.643$^{+0.065}_{-0.069}$	&	6.09$^{+0.041}_{-0.043}$	&	0.794$^{+0.073}_{-0.075}$ \\
GC0041	&	0.296$^{+0.094}_{-0.047}$	&	4.82$^{+0.054}_{-0.028}$	&	0.711$^{+0.079}_{-0.043}$ \\
GC0040	&	0.026$^{+0.094}_{-0.040}$	&	4.97$^{+0.056}_{-0.026}$	&	0.593$^{+0.085}_{-0.049}$ \\
GC0227	&	0.098$^{+0.063}_{-0.067}$	&	4.72$^{+0.037}_{-0.040}$	&	0.682$^{+0.057}_{-0.060}$ \\
\enddata
\tablecomments{$\hat{\kappa}$-space parameters as derived from Eqs. (9), (10) and (11).}
\end{deluxetable}

\end{document}